\documentclass{aa}
\usepackage{txfonts}
\usepackage{graphicx} 
\usepackage{natbib}
\usepackage{threeparttable}
\bibpunct{(}{)}{;}{a}{}{,}

\def \hcm {\hbox {\ifmmode $ atom cm$^{-2}\else atom cm$^{-2}$\fi}}
\def \arcmin {\hbox{$^\prime$}}
\def \arcsec {\hbox{$^{\prime\prime}$}}
\def \deg {$^{\circ}$}

\def \approxgt{\mathrel{\hbox{\rlap{\lower.55ex \hbox {$\sim$}}
        \kern-.3em \raise.4ex \hbox{$>$}}}}
\def \approxlt{\mathrel{\hbox{\rlap{\lower.55ex \hbox {$\sim$}}
        \kern-.3em \raise.4ex \hbox{$<$}}}}

\newcommand {\Msun}{M_\odot}

\def\ang {$\rm\AA$}

\def \grs {GRS\,1915+105}

\def \grso {GRS\, 1758--258}
\def \oe {1E 1740.7--2942}
\def \gx {GX\,13+1}

\def \scox {Sco\,X$-$1}
\def \cirx {Cir\,X$-$1}
\def \cygx {Cyg\,X$-$1}

\begin{document}

\title{A search for signatures of interactions of X-ray binary outflows with their environments with ALMA}

\author{
M. D{\'i}az Trigo\inst{1} \and D. Petry\inst{1} \and E. Humphreys\inst{1,2} \and C. M. V. Impellizzeri\inst{3,4} \and H. B. Liu\inst{5}
}
\institute{
ESO, Karl-Schwarzschild-Strasse 2, D-85748 Garching bei M\"unchen, Germany
\and
Joint ALMA Observatory, Av. Alonso de Cordova 3107, Vitacura, Santiago, Chile
\and
Leiden Observatory, Leiden University, PO Box 9513, 2300 RA Leiden, The Netherlands 
\and
National Radio Astronomy Observatory, Charlottesville, VA, USA
\and
Academia Sinica Institute of Astronomy and Astrophysics, No. 1, Section 4, Roosevelt Road, Taipei 10617, Taiwan, Republic of China
}

\date{Received ; Accepted:}

\authorrunning{D{\'i}az Trigo et al.}

\titlerunning{Interactions of X-ray binary outflows with their environments}

\abstract{We observed the X-ray binaries \cirx, \scox, \grs, \gx,\ and \cygx\ with the Atacama Large Millimeter/submillimeter Array (ALMA). Unresolved continuum emission is found at the positions of all the sources at a frequency of 92~GHz, with flux densities ranging between 0.8 and 10 mJy/beam. In all cases the emission can be associated with jets that have been extensively observed at lower frequencies.
We searched for line emission from H$\alpha$ recombination, SiO, H$_2$O, and CH$_3$OH at the positions of all the sources and, for \cirx\ and \cygx, also at regions where shocks associated with an interaction between the jet and the interstellar medium had previously been observed. The search did not yield any significant detection, resulting in 3$\sigma$ upper limits between 0.65 and 3.7~K~km~s$^{-1}$ for the existence of line emission in these regions. 

In contrast, we detected spatially unresolved SiO emission in the field of view of \gx,\ and we tentatively associate this emission with a SiO maser in a potential young stellar object or evolved star. We also found spatially extended line emission at two additional sites in the field of view of \gx\ that we tentatively associate with emission from SO and CH$_3$OH; we speculate that it may be associated with a star-forming region, but again we cannot rule out alternative origins such as emission from evolved stars. }
  
\maketitle

\section{Introduction}
\label{sec:intro}

Accretion onto stellar-mass black holes (BHs) and neutron stars (NSs) in X-ray binaries (XRBs) releases copious amounts of radiation into the interstellar medium \citep[ISM;][]{frank92apia}. X-ray binaries also show the two other fundamental ingredients of accretion onto compact objects: relativistic jets \citep{cygx1:stirling01mnras,scox1:fomalont01apj} and hot, dense winds \citep{gx13:ueda04apj, 1655:miller06nat, v404:munoz-darias16nature}. 

Given the similarity between accretion processes in XRBs and other compact objects, such as supermassive BHs, where radiation and outflows interact with their host galaxy \citep[see e.g. reviews by][]{mcnamara07araa,fabian12araa} and even regulate galaxy growth \citep[e.g.][]{maiolino17nat}, we also expect the XRB environments to be affected by the expelled material, energy, and momentum. As an example, jets are likely at the origin of the parsec-scale bubbles observed around ultra-luminous X-ray sources \citep[e.g.][]{pakull10nature} that are generally identified with (extragalactic) XRBs containing NSs or stellar-mass BHs and accreting well above the Eddington limit, or, in a few cases, intermediate-mass BHs accreting at the Eddington limit. However, in our galaxy, and despite extensive searches for signatures of an interaction between XRB jets and the ISM \citep[and references therein]{miller-jones08aip}, direct observational evidence of jet-inflated bubbles has only been reported in two cases \citep{ss433:dubner98aj,cygx1:gallo05nature,cygx1:russell07mnras}; one of these two cases, SS~433, is the only known galactic ultra-luminous X-ray source \citep{ss433:begelman06mnras}. 

Recently, and following previous work by \citet{rodriguez98aa} and \citet{1915:chaty01aa}, \citet{1915:tetarenko18mnras} searched for molecular line emission in the IRAS 19132+1035 region near the microquasar \grs. They identified a zone with significant molecular line emission that could be traced back to a jet-ISM interaction but concluded that the jet did not appear to be powering the entire region. \citet[][]{tetarenko20mnras} also reported a detection of molecular line emission in the central molecular zone coinciding with candidate jet-ISM interaction zones near the BH XRBs \grso\ and \oe\ that may trace the cavities in the gas surrounding those systems. Importantly, all these works report on jet-ISM interactions taking place relatively far from the XRBs, at parsec scales.  

At smaller scales (a few tens of AU), \citet{scox1:fomalont01apj} detected radio lobes dominated by compact hot spots along the jet direction of the NS XRB \scox; they speculated that the radio lobes were generated at the working surface where the highly relativistic jet impacted the ambient medium. At even smaller scales, the effect of wide-angle X-ray winds from the accretion disc could become noticeable. Being dense ($\sim$10$^{12}$--10$^{13}$ cm$^{-3}$), relatively slow (with velocities of a few hundred km~s$^{-1}$), and with estimated mass loss rates of 10$^{18}$--10$^{19}$~g~s$^{-1}$\citep[e.g.][]{gx13:ueda04apj,1630:kubota07pasj,1915:ueda09apj}, such winds could have a significant impact close to the XRBs by sweeping up the ambient gas and compressing it. Alternatively, winds could cool, form cold clumps, and result in molecular outflows, at least in luminous quasars \citep[][and references therein]
{richings18mnras}. However, to our knowledge, the effect of XRB winds in the ISM or their evolution as they leave the binary system has not yet been investigated. 

While in terms of energetics XRB winds have been estimated to be less powerful than jets \citep[10$^{33}$--10$^{35}$~erg s$^{-1}$ vs. $\sim$\,10$^{36}$ --10$^{37}$~erg s$^{-1}$, respectively; e.g. ][]{king13apj, fender16lnp}, the wind power estimates suffer uncertainties related to the wind optical thickness, the spectral energy distribution that is illuminating it, and the portion of the wind that is invisible via X-ray spectroscopy due to being fully ionised \citep[e.g.][]{diaz16an}. Moreover, the duty cycle of winds is still uncertain, the question being whether they co-exist with jets throughout an accretion outburst \citep[e.g.][]{1915:neilsen09nat,homan16apj}. If so, their duty cycle would be longer than that of jets, making the total energy deposited over time more similar for both output channels.
Finally, the high collimation and speed of jets may make a significant energy deposit in their environments unlikely until they encounter a dense medium \citep{heinz02aa,miller-jones08aip} or are significantly decelerated \citep{1550:corbel02science}. In contrast, accretion disc winds may be more efficient at depositing energy in low density environments at sub-parsec scales due to their lower velocity and wide-angle geometry.  In the end, the relative impact of jets, winds, and radiation will not only depend on their properties but also on the properties of the ISM where the XRBs reside. For example, \citet{haid18mnras} found that stellar winds couple more efficiently to the ISM than ionising radiation but that the relative impact of those winds and the ionising radiation is strongly dependent on the properties of the ambient medium. 

We observed five XRBs at millimetre wavelengths with the Atacama Large Millimeter/submillimeter Array (ALMA) with the goal of exploring the existence of atomic and molecular line emission from their accretion discs, both close to the XRBs, where powerful X-ray and optical winds are expelled, and farther away, where there are known sites of jet-ISM interaction and where shocks are expected.
For this exploratory study we chose a sample of persistently accreting sources that includes both NS and BH systems and both low- and high-mass stellar companions. All sources are known jet emitters, and strong wide-angle winds have been well studied for a subset of them. We targeted molecular lines that trace high density gas commonly found in shocks and atomic recombination lines. In this paper, we report on the results of these observations. 
\section{Characteristics of the sample}
\subsection{\cirx}

\cirx\ is the youngest known XRB, with an estimated age of $<$5400~years \citep{cirx1:heinz13apj}. It is located at a distance of 9.4\,$^{+0.8}_{-1.0}$~kpc \citep{cirx1:heinz15apj}. It contains a star, the type of which remains in dispute \citep[e.g.][]{cirx1:jonker07mnras,cirx1:johnston16mnras}, that orbits a NS \citep{cirx1:tennant86mnras,linares10apj} in an eccentric orbit ($e \sim$\,0.4--0.45).  
Periodic flares at radio, infrared (IR), and X-ray frequencies every 16.6~d were interpreted as enhanced accretion close to periastron \citep{cirx1:murdin80aa,cirx1:nicolson80mnras}. 
Radio observations at 5.5 and 9~GHz covering the full binary orbit showed that significant emission is present throughout the orbit, although the flux level is very low, $\approxlt$1~mJy, at phases outside periastron \citep{cirx1:calvelo12mnrasb}. The same observations showed that the spectra evolve from optically thick to thin during the course of a flare, reaching already optical thinness by orbital phases of 0.05. 

\citet{cirx1:calvelo12mnras} reported, for the first time, millimetre emission (at 33 and 35 GHz) during a periastron flare that had structures at sub-arcsecond scales (1\arcsec\ corresponding to 0.04~pc at the distance of \cirx) and decayed over the following four days. A comparison of the structure of this emission with the radio emission resolved at milliarcsecond scales by \citet{miller-jones12mnras} revealed a change in the position angle of the elongated structure, which has a size of 185~AU and had been identified as the jet, suggesting that it could be precessing. Indeed, \citet{cirx1:coriat19mnras} successfully modelled the emission observed at different frequencies and angular scales from sub-arcsecond to arcminutes as a jet precessing over a five-year period and with mildly relativistic ejecta.
Jets have also been observed in X-rays and are spatially coincident with the radio jets \citep{cirx1:heinz07apj,cirx1:sell10apj}. The same X-ray observations show two diffuse X-ray caps that could be terminal shocks resulting from powerful jets running into the ISM \citep{cirx1:sell10apj}. 
Besides the collimated jets, an accretion disc wind likely driven by thermal and radiation pressure has been inferred from the detection of X-ray P--Cygni profiles \citep{cirx1:brandt00apjl,cirx1:schulz02apj}, and a stellar wind from the companion has been inferred from the detection of blueshifted X-ray emission lines at a different epoch \citep{cirx1:schulz20apj}.

\cirx\ is embedded in a radio nebula that extends over several parsecs \citep{haynes86nat} and is attributed to the supernova remnant at which \cirx\ originated \citep{cirx1:heinz13apj}. Although not at the origin of the nebula as initially suggested \citep{cirx1:stewart93mnras}, there is increasing evidence that the precessing jets of \cirx\ may be shaping it by injecting energy at different scales and in different directions over its five-year cycle \citep{cirx1:coriat19mnras}.

\subsection{\scox}

At a distance of 2.8\,$\pm$\,0.3 kpc  \citep{bradshaw99apjl}, \scox\ is the closest known `persistent' NS XRB. It constantly radiates at or near Eddington luminosity and has a period of 0.787 days \citep{gottlieb75apj}.  \citet{steeghs02apj} performed high-resolution IR spectroscopy throughout 75\% of the orbit of \scox\ and suggested that the companion star must be a significantly evolved subgiant with a radius of more than twice the main-sequence radius of 0.5\,R$_\odot$ and a mass of $\sim$\,0.42\,M$_\odot$. Radio emission from the source was first detected by \citet{andrew68nat}. Measurements with the Very Large Baseline Array (VLBA) spanning three years 
revealed the existence of two components at $\sim$20--40~milliarcseconds of the radio core (corresponding to $\sim$20--170~AU at the distance of \scox), which are moving away from it and are reminiscent of the hot spots found in many extragalactic radio double sources \citep{scox1:fomalont01apj}.
Their relative motion and flux densities implied an average component speed of $v/c$\,=\,0.45\,$\pm$\,0.03 at 44$\pm$6$^{\circ}$\ to the line of sight. \citet{scox1:fomalont01apj} suggested that the two radio components consisted of ultra-relativistic plasma produced at a working surface where the highly relativistic beams impacted the ambient medium. 

Evidence of a high-velocity outflow was also reported at near-infrared (NIR) frequencies based on the detection of a P-Cygni profile in the Brackett $\gamma$ line of \scox\ \citep{gx13:bandyo99mnras}. The existence of the Brackett $\gamma$ line was later confirmed, and its width was measured to be $\sim$\,480~km s$^{-1}$, consistent with that of H\,$\alpha$ \citep{scox1:matasanchez15mnras}, although the P-Cygni profile indicative of a wind was not confirmed.  

\subsection{\grs}

\grs\ is a low-mass XRB (LMXB) at a distance of 8.6~kpc \citep{1915:reid14apj} that consists of a K$_{III}$ companion \citep{1915:greiner01aa} orbiting a 12~M$_{\odot}$ BH with a period of 33.85\,$\pm$\,0.16~days \citep{1915:steeghs13apj}. It was the first XRB to show episodic radio jets with apparently superluminal velocities \citep{1915:mirabel94nature}.  Besides the episodic jets, a compact radio jet of 2.5--7~AU was resolved in observations at 15--43~GHz \citep{1915:dhawan00apj}. 

\citet{rodriguez98aa} first studied the surroundings of \grs\ and investigated the possible relation of the XRB with
two IR-to-radio sources that appear symmetrically located
with respect to \grs\ and aligned with the position
angle of the relativistic ejecta. They reported on a potential non-thermal structure in one of these regions, IRAS~19132+1035, that could be related to either a bow shock or an ionisation front. \citet{1915:chaty01aa} explored these regions further and reported that the evidence for the non-thermal structure being a jet-ISM interaction zone was indecisive. A similar conclusion was reached by \citet{miller-jones08aip}, who set upper limits to the presence of X-ray emission that could signal the presence of shock acceleration. Finally, \citet{1915:tetarenko18mnras} re-observed the zone of significant molecular line emission and concluded instead that it could be traced back to a jet-ISM interaction but also that the jet did not appear to be powering the entire region. 

Similarly to other LMXBs observed at high inclination, \grs\ also shows an accretion disc wind, observed via blueshifted absorption lines in X-ray spectra \citep[e.g.][]{1915:kotani00apj,1915:lee02apj,1915:ueda09apj}. The wind is generally visible whenever the jet is absent \citep{1915:neilsen09nat} and has been shown to react to changes in the broadband X-ray spectrum on timescales as short as 5 seconds \citep{1915:neilsen11apj}.    
\subsection{\gx}

\gx\ is a NS LMXB \citep{gx13:fleischman85aa, gx13:matsuba95pasj} observed at a high inclination \citep{gx13:diaz12aa,gx13:iaria14aa}. The companion star was identified as an evolved late-type K5~{\sc iii} star, located at a
distance of 7$\pm$1~kpc \citep{gx13:bandyo99mnras}. An orbital period of 24.7~days was determined from IR and X-ray modulations \citep{gx13:bandyopadhyay02apj,gx13:corbet03apj,gx13:corbet10apj} and later refined to 24.5274(2)~days based on the timing of X-ray dips \citep{gx13:iaria14aa}. Radio emission at centimetre wavelengths was first detected by \citet{gx13:grindlay86apj}. The similarity of the radio flux levels, spectral indices, and their variability with those of other NSs accreting at high Eddington luminosities was interpreted as an indication of the presence of relativistic jets \citep[e.g.][]{gx13:garcia88apj,gx13:homan04aa}. 

In addition to jets, \gx\ shows narrow absorption lines in the X-ray band that have been interpreted as a wide-angle accretion disc wind outflowing at $\sim$ 400 km~s$^{-1}$ \citep{gx13:ueda01apjl, gx13:sidoli02aa, gx13:ueda04apj,gx13:diaz12aa,gx13:madej14mnras,gx13:allen18apj}. Evidence for a disc wind was also reported at NIR frequencies based on the presence of a Brackett $\gamma$ line with a P-Cygni profile that indicates an outflow velocity of $\sim$2400 km~s$^{-1}$ \citep{gx13:bandyo99mnras}. 

The existence of an extended nebula surrounding the source similar to those of \cirx\ \citep{haynes86nat} and \cygx\ \citep{cygx1:gallo05nature} has not been reported so far, and \citet{gx13:garcia88apj} attribute extended emission found in the area around \gx\ to unassociated galactic H$_{II}$ regions. 
They also set a 2$\sigma$ upper limit of 0.1~mJy at 6~cm and 1.0~mJy at 20~cm to the existence of \scox-like lobes within 1\arcmin\ of \gx\ (equivalent to $\sim$2~pc at a distance of 7~kpc).

\subsection{\cygx}

\cygx\ is one of the best studied BH XRBs and the only confirmed high-mass XRB (HMXB) in our sample. The binary has a period of 5.6~days \citep{cygx1:holt79apj}, was most likely born in the stellar association Cygnus OB3 \citep{cygx1:mirabel03science}, and consists of a BH accreting from the stellar wind of a supergiant O9.7 star \citep{cygx1:webster72nat,cygx1:walborn73apj}. At a distance of 2.22\,$^{+0.18}_{-0.17}$~kpc, according to the most recent estimate with VLBA \citep{miller-jones21science}, the BH would have a mass of 21.2~M$_{\odot}$. \cygx\ is one of the few persistently accreting BHs (\grs\ is another) and is predominantly in a spectrally hard accretion state \citep[e.g.][]{grinberg13aa}. During such a state, the spectrum between 2 and 220~GHz shows a flat spectral index \citep{cygx1:fender00mnras} and is attributed to a compact jet, resolved at milliarcsecond scales \citep[$\sim$33~AU at the distance of \cygx; ][]{cygx1:stirling01mnras}. 
The line-driven stellar wind, focused towards the BH \citep[e.g.][]{cygx1:gies03apj}, is clumpy \citep[e.g.][]{cygx1:rahoui11apj,cygx1:hirsch19aa}, dominates mass transfer, and is in turn affected by the X-rays of the BH \citep[e.g.][]{cygx1:gies03apj,cygx1:gies08apj,cygx1:nowak11apj}. 

\cygx\ is the only source in our sample associated with a parsec-scale nebula detected in radio \citep{cygx1:gallo05nature} and bright in H$\alpha$ and [O$_{III}$] (5007 \ang) emission \citep{cygx1:russell07mnras} that is thought to be inflated by the BH jet or the wind of the O star \citep{sell15mnras}. Remarkably, the energy output of the O star wind is estimated to be of the order of that of the BH jet, indicating that, in addition to the two feedback channels considered here for XRBs (jets and accretion disc winds), stellar winds also have to be considered for HMXBs (we note that stellar winds from massive stars continue to be investigated as a source of feedback, and some authors even consider them to be more disruptive for molecular clouds than supernovae, e.g. \citet[][]{reyraposo2017mnras}; however, these investigations often do not consider the role of the compact object in shaping the stellar wind). Finally, the nebula around \cygx\ has an age similar to the time that the XRB was close to a bright H$_{II}$ region, indicating that a dense local medium may be required to form the shock wave \citep{cygx1:russell07mnras}. 
  
\section{Observations and data analysis} 
\label{sec:observations}

\begin{table*}
\begin{center}
\begin{threeparttable}[b]
\caption[]{Observation log. LMXB and HMXB correspond to low-mass and high-mass X-ray binaries, respectively. NS and BH indicate the nature of the compact object in the X-ray binary, neutron star or black hole, respectively. We note that \cygx\ had to be observed on two different days to reach the requested sensitivity.}
\begin{tabular}{l@{\extracolsep{1mm}}cccccc@{\extracolsep{1mm}}}
\hline \noalign {\smallskip}
Source & Type & Compact  & Orbital & Distance & Inclination & Observation date \\
& & object & period (d) & (kpc) & (degrees) \\
\hline \noalign {\smallskip}
\cirx\ & HMXB/LMXB?\tnote{1} & NS\tnote{2} & 16.6\tnote{3} & 9.4\tnote{4} & $\approxgt$70\tnote{5} & 23 Oct 2016 17:26 -- 18:45 \\
\scox\ & LMXB\tnote{6} & NS\tnote{6} & 0.79\tnote{7} & 2.8\tnote{8} & $\approxlt$40\tnote{9} & 25 Oct 2016 16:50 -- 17:38 \\
\grs\ & LMXB\tnote{10} & BH\tnote{11} & 33.8\tnote{12} & 8.6\tnote{13} & 66\,$\pm$\,2\tnote{11}  & 25 Oct 2016 17:41 -- 18:23 \\ 
\gx\ & LMXB\tnote{14} & NS\tnote{14} & 24.5\tnote{15} & 7\tnote{16}& $\approxgt$70\tnote{17} & 05 Nov 2016 22:46 -- 23:27 \\
\cygx\ & HMXB\tnote{18} & BH\tnote{19} & 5.6\tnote{20} & 2.22\tnote{21}&  27.1\,$\pm$\,0.8\tnote{19} & 06 Nov 2016 21:02 -- 22:02 \\
& & & & & & 29 Nov 2016 20:39 -- 21:40 \\
\noalign {\smallskip} \hline 
\label{tab:obslog}
\end{tabular}
\begin{tablenotes}[para,flushleft]\footnotesize
\item[1] \citet{cirx1:jonker07mnras,cirx1:johnston16mnras}
\item[2] \citet{cirx1:tennant86mnras,linares10apj} 
\item[3] \citet{cirx1:kaluzienski76apjl}
     \item[4] \citet{cirx1:heinz15apj}  
     \item[5] From the presence of dips in its light curve \citep{cirx1:clarkson04mnras}
     \item[6] \citet{steeghs02apj}      
     \item[7]\citet{gottlieb75apj}
     \item[8]\citet{bradshaw99apjl} 
     \item[9] \citet{scox1:matasanchez15mnras}
     \item[10]\citet{1915:greiner01aa}
     \item[11]\citet{greiner01nat}
     \item[12]\citet{1915:steeghs13apj}
     \item[13] \citet{1915:reid14apj} 
     \item[14]\citet{gx13:fleischman85aa, gx13:matsuba95pasj}
     \item[15] \citet{gx13:bandyopadhyay02apj}
     \item[16]\citet{gx13:bandyo99mnras} 
     \item[17] From the presence of dips in its light curve \citep{gx13:diaz12aa,gx13:iaria14aa}
     \item[18]\citet{cygx1:webster72nat}
     \item[19] \citet{cygx1:orosz11apj}
     \item[20]\citet{cygx1:holt79apj} 
     \item[21]\citet{miller-jones21science}

\end{tablenotes}
\end{threeparttable}
\end{center} 
\end{table*}

\begin{figure}[ht]
\includegraphics[angle=0.0,width=0.5\textwidth]{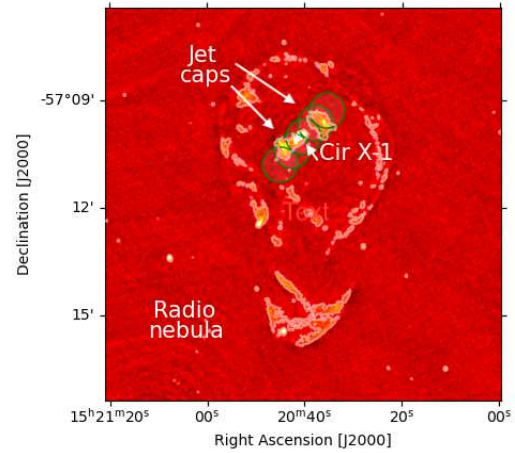}
\caption{ALMA pointings along the jet direction of \cirx\ (in green) overlaid on the radio image from \citet{cirx1:coriat19mnras}.}
\label{fig:cirx_atca}
\end{figure}

\begin{figure}[ht]
\hspace{0.2cm}
\includegraphics[angle=0.0,width=0.48\textwidth]{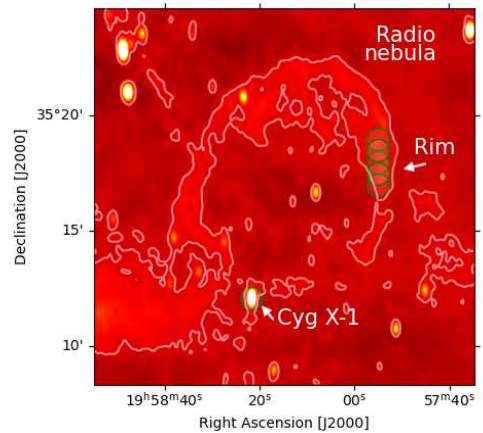}\caption{ALMA pointings on \cygx\ and the shell of the nebula surrounding \cygx\ (in green) overlaid on the radio image from \citet{cygx1:gallo05nature}.}
\label{fig:cygx_gallo}
\end{figure}

ALMA observed the XRBs \cirx, \scox, \grs, \gx,\ and \cygx\ from October to November 2016 (see Table~\ref{tab:obslog}). For these exploratory observations we only used the 12 m array due to its superior sensitivity and angular resolution, which was a better match for our search for weak and small regions of line emission.
All observations were set up to use six spectral windows centred at 84.68117, 85.64439, 86.24337, 86.84696, 96.49964, 97.5, and 99.02295~GHz (Band~3). The characteristics of the spectral windows are listed in Table~\ref{tab:spwlog}. The observations were performed using a number of 12 m antennas, between 37 and 44, and maximum baseline lengths between 0.7 and 1.8 km. One pointing was performed in the direction of each source, and we performed additional pointings for \cirx\ along the jet direction (a total of five pointings extending over a total of 189\arcsec$\times$63\arcsec\ centred on the source; see Fig.~\ref{fig:cirx_atca}) and for \cygx\ at the shell of the nebula surrounding it (a total of five pointings extending over 189\arcsec$\times$63\arcsec\ and centred at position 19:57:55.0 +35:18:30.0; see Fig.~\ref{fig:cygx_gallo}). Further technical characteristics of the observations are listed in Table~\ref{tab:obslog2}.

We first examined the ALMA calibrated products. Re-calibration for all datasets as well as image generation and analysis was done with CASA version 5.6.1-8 \citep{emonts19casa} and the integrated ALMA pipeline. We re-generated images from the calibrated products following standard procedures and applying natural weighting. We then applied phase-only self-calibration with solution intervals of 15, 90, 90, and 45~s for \cirx, \scox, \grs,\ and \cygx, respectively. We did not apply self-calibration to \gx\ as we deemed it too faint. 
We found that self-calibration significantly improved the images, recovering $\sim$\,30--45\% of the flux at the source position that was otherwise spread outside of one synthesised beam due to phase decoherence. We attribute this phase decoherence to the relatively poor ($\approxgt$ 30\deg) phase rms conditions during all the observations except for \gx\ (see Table~\ref{tab:obslog2}). While ALMA antennas are equipped with water vapour radiometers, which are used to correct differential phase errors introduced by precipitable water vapour (PWV) over the array, this correction is less efficient in conditions of very low PWV, as is the case for our observations (for reference, the maximum PWV value required for Band 3 observations is 5.186 mm\footnote[1]{Remijan, A., Biggs, A., Cortes, P., A., Dent, B., Di Francesco, J., Fomalont, E., Hales, A., Kameno, S.,
Mason, B., Philips, N., Saini, K., Stoehr, F., Vila Vilaro, B., Villard, E. 2019, ALMA Technical Handbook,
ALMA Doc. 7.3, ver. 1.1}). 

We also extracted data cubes at the spectral resolution of the cubes, $\approxlt$1~km s$^{-1}$, and inspected them for the presence of line emission. 
Unless otherwise indicated, the reported flux density for all sources in this paper corresponds to the peak flux from the elliptical Gaussian fitted to the source emission in the image plane after primary beam correction, and the error corresponds to the rms of the image measured far away from the source position.

\begin{table*}
\begin{center}
\caption[]{Correlator setup.}
\begin{scriptsize}
\begin{tabular}{llllll}
\hline \noalign {\smallskip}
Line & Rest frequency & Bandwidth & Resolution & Spw \\
& (GHz) & (MHz) & (MHz) \\
\hline \noalign {\smallskip}
Upper Side Band \\
\noalign {\smallskip}
H (40) $\alpha$ & 99.02295 & 937.5 & 0.488 & 0 \\
Continuum & 97.495 & 468.75 & 0.488 & 1 \\
H$_2$O v2=1 4(4,0)-5(3,3) & 96.49964 & 468.75 & 0.488 & 2 \\
\hline \noalign {\smallskip}
Lower Side Band \\
\noalign {\smallskip}
SiO v=1 2-1 & 86.24337 & 468.75 & 0.488 & 3 \\
SiO v=0 2-1 & 86.84696 & 468.75 & 0.488 & 4 \\
CH$_3$OH & 84.68628 & 468.75 & 0.488 & 5 \\
H (42) $\alpha$ & 85.64956 & 468.75 & 0.488 & 6 \\
\noalign {\smallskip} 
\noalign {\smallskip} \hline 
\label{tab:spwlog}
\end{tabular}
\end{scriptsize}
\end{center} 
\end{table*}

\begin{table*}
\begin{center}
\caption[]{Technical characteristics of the ALMA observations. ToS is the time on source in seconds. FoV is the field of view of the observations; it corresponds to a single pointing centred on the XRB, except for the case of \cirx,\ for which five consecutive pointings overlapping with Nyquist frequency and centred on the source were performed (we took as reference the primary beam of the ALMA antennas at 92~GHz, 63\arcsec). The synthesised beam is the beam achieved in the images. For the calculation of the physical scales in parsecs, we used the distances of the sources listed in Table~\ref{tab:obslog}. PWV is the average precipitable water vapour throughout each observation. The phase rms is shown before and after correcting differential phase errors introduced by PWV over the array. For \cygx,\ the PWV and the phase rms are an average of the values of the two performed observations.
}
\begin{scriptsize}
\begin{tabular}{l@{\extracolsep{1.2mm}}ll@{\extracolsep{1.1mm}}l@{\extracolsep{1mm}}ll@{\extracolsep{1.2mm}}ll@{\extracolsep{1.2mm}}lll}
\hline \noalign {\smallskip}
Source & ToS  & \multicolumn{3}{c}{Calibrators} & \multicolumn{2}{c}{FoV} & \multicolumn{2}{c}{Synthesised beam} & PWV &  Phase rms \\
& (s) & Bandpass & Flux & Phase & (arcsec) & (parsec) & (arcsec) & (parsec) & (mm) & before/after (\deg) \\
\hline \noalign {\smallskip}
\cirx\ & 4768 & J1617$-$5848 & Titan & J1531$-$5108 & 63\,$\times$\,189 & 2.87\,$\times$\,8.61 & 0.67\,$\times$\,0.63 & 0.030\,$\times$\,0.029 & 1.5 &63/29 \\
\scox\ & 2928 & J1517$-$2422 & Callisto & J1629$-$1720 & 63\,$\times$\,63 & 0.86\,$\times$\,0.86 & 0.78\,$\times$\,0.65 & 0.011\,$\times$\,0.009 & 0.6 & 54/46 \\
\grs\ & 2527 & J1751+0939 & J1751+0939 & J1922+1530 & 63\,$\times$\,63 & 2.63\,$\times$\,2.63 & 1.44\,$\times$\,0.63  & 0.060\,$\times$\,0.026  & 0.5 & 32/24 \\
\gx\ & 2421 & J1924$-$2914 & J1924$-$2914 & J1832$-$2039 & 63\,$\times$\,63 & 2.14\,$\times$\,2.14 & 1.34\,$\times$\,0.80 & 0.045\,$\times$\,0.027 & 1.7 & 14/10 \\
\cygx\ & 7249 & J2025+3343 & J2148+0657 & J1953+3537 & 63\,$\times$\,63 & 0.57\,$\times$\,0.57 & 1.65\,$\times$\,0.96 & 0.015\,$\times$\,0.009 & 2.9 & 44/34 \\
\noalign {\smallskip}
\noalign {\smallskip} \hline 
\label{tab:obslog2}
\end{tabular}
\end{scriptsize}
\end{center} 
\end{table*}

\begin{figure*}[ht]
\hspace{-1cm}
\includegraphics[angle=0.0,width=1.1\textwidth]{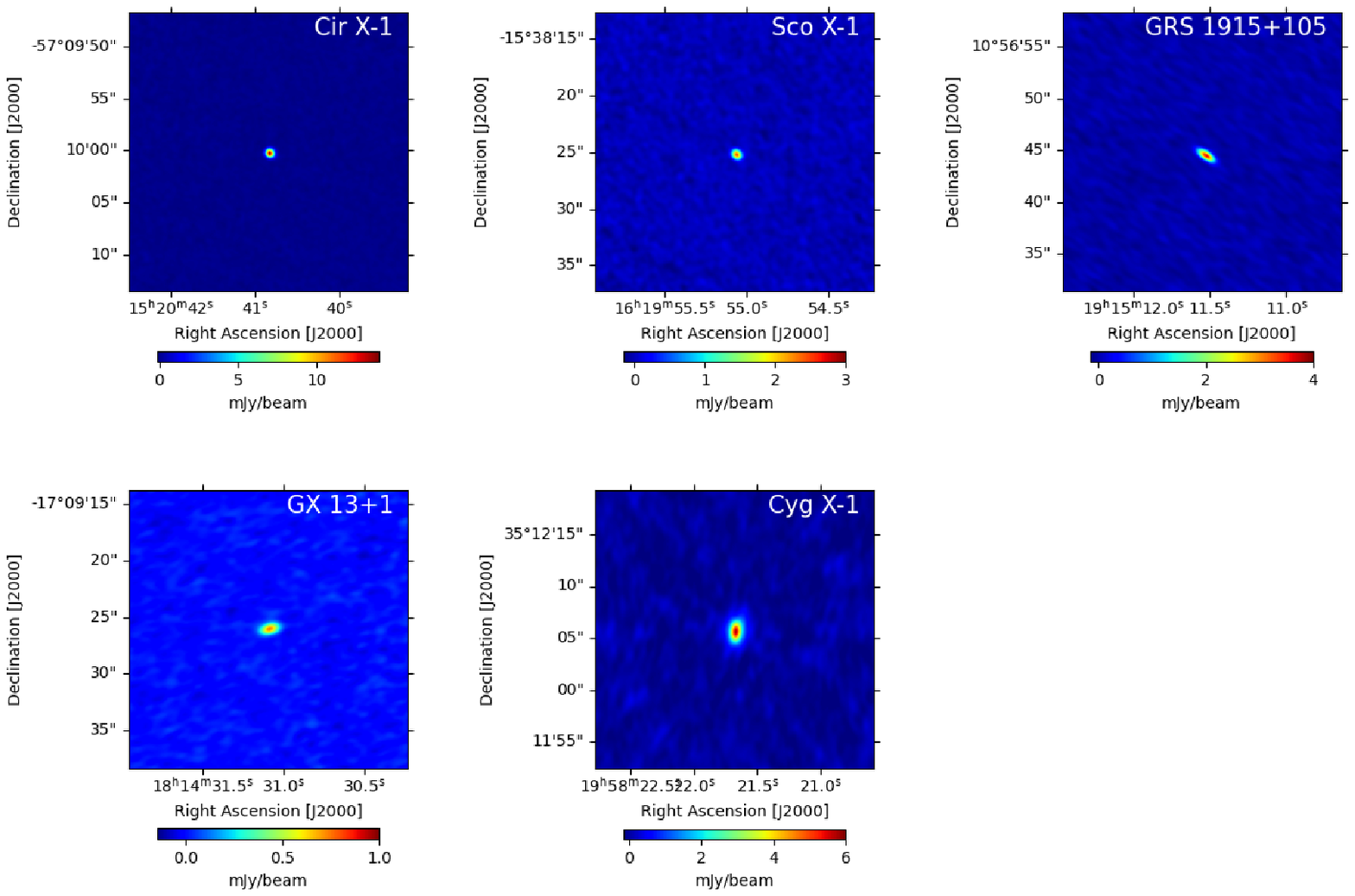}
\vspace{-1cm}
\caption{ALMA images of the averaged continuum of \cirx, \scox, \grs, \gx,\ and \cygx. 
The synthesised beams are listed in Table~\ref{tab:obslog}.}
\label{fig:cont_images}
\end{figure*}

\section{Results}

We detected significant spatially unresolved continuum emission at the position of all the XRBs in our sample (see Fig.~\ref{fig:cont_images} and Table~\ref{tab:fluxlog}). The flux densities range from 0.7 to 13.7~mJy/beam, and the spectral index between 84 and 99 GHz is flat or inverted for all the sources except for \cirx.  

\begin{figure*}[ht]
\includegraphics[angle=0.0,width=.95\textwidth]{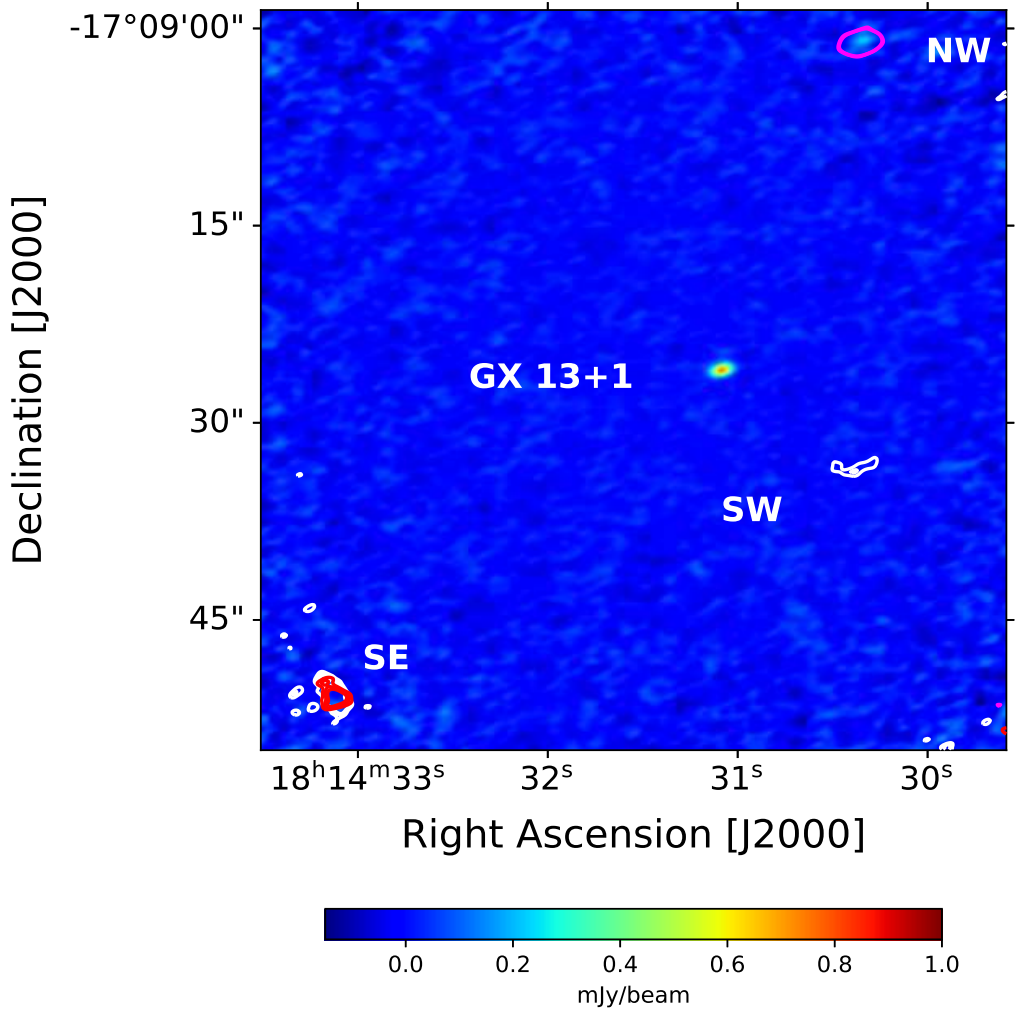}
\caption{Continuum emission (coloured image) associated with the relativistic jet at the position of \gx. The contours show the integrated intensity maps at the frequency of the lines detected around \gx\ (see Fig.~\ref{fig:line_images} and Table~\ref{tab:lineloggx} for details). The emission lines at each position are shown in Fig.~\ref{fig:line_images}. Magenta, red, and white contours correspond to the lines at 86.23, 96.73, and 99.29~GHz, respectively. The excess in the average continuum map of \gx\ at the NW position is due exclusively to the line emission at that position.}\label{fig:gx13_composite}
\end{figure*}

\begin{figure*}[ht]
\includegraphics[angle=0.0,width=0.95\textwidth]{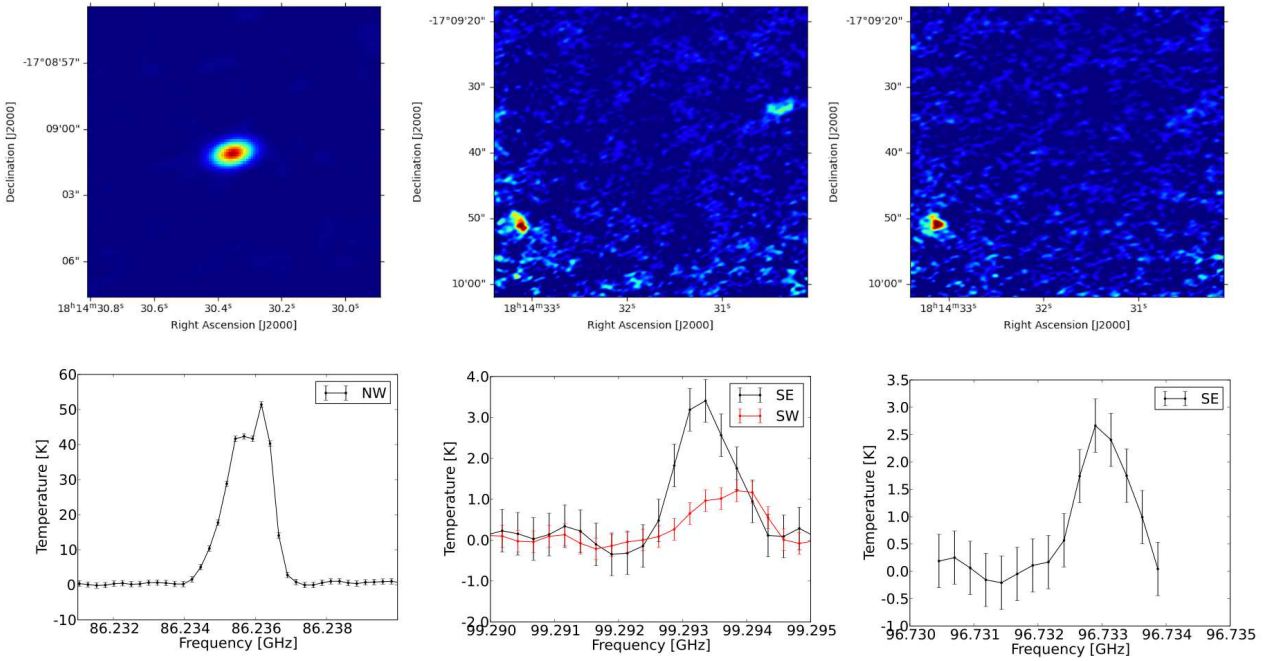}
\caption{Detail of regions showing line emission in the field of view of \gx. Upper panels: Integrated intensity maps after primary beam correction within the frequency range indicated in the first column of Table~\ref{tab:lineloggx}. Lower panels: Emission line in the spatial region of emission derived from the integrated intensity maps.}
\label{fig:line_images}
\end{figure*}

\begin{figure*}[ht]
\includegraphics[angle=0.0,width=0.95\textwidth]{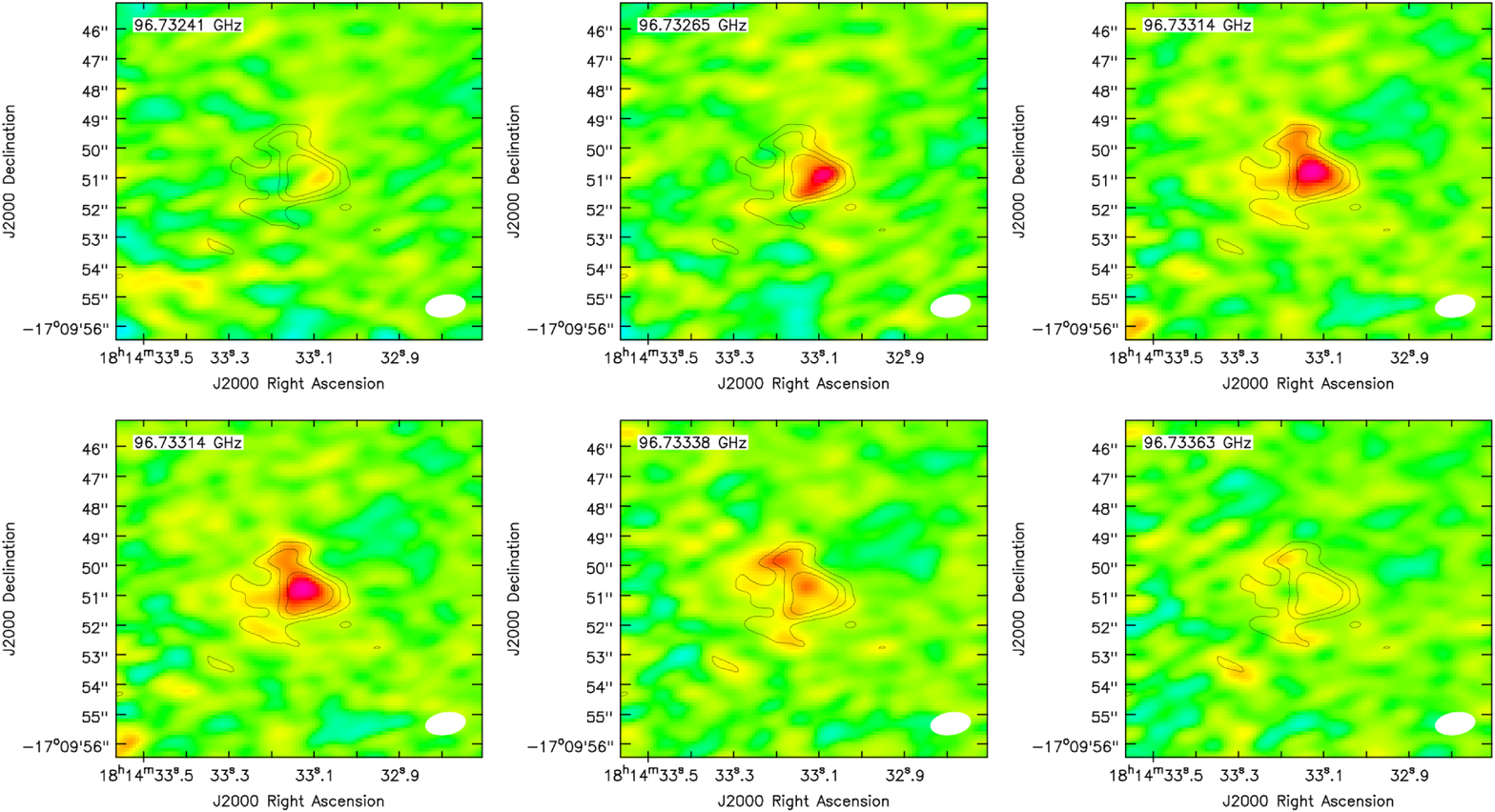}
\caption{Detail of regions showing line emission in the field of view of \gx: channel map at the SE position for the line centred at 96.7329~GHz. The contours show the integrated intensity after primary beam correction within the frequency range indicated in the first column of Table~\ref{tab:lineloggx}.
}
\label{fig:line_maps1}
\end{figure*}

\begin{figure*}[ht]
\includegraphics[angle=0.0,width=0.95\textwidth]{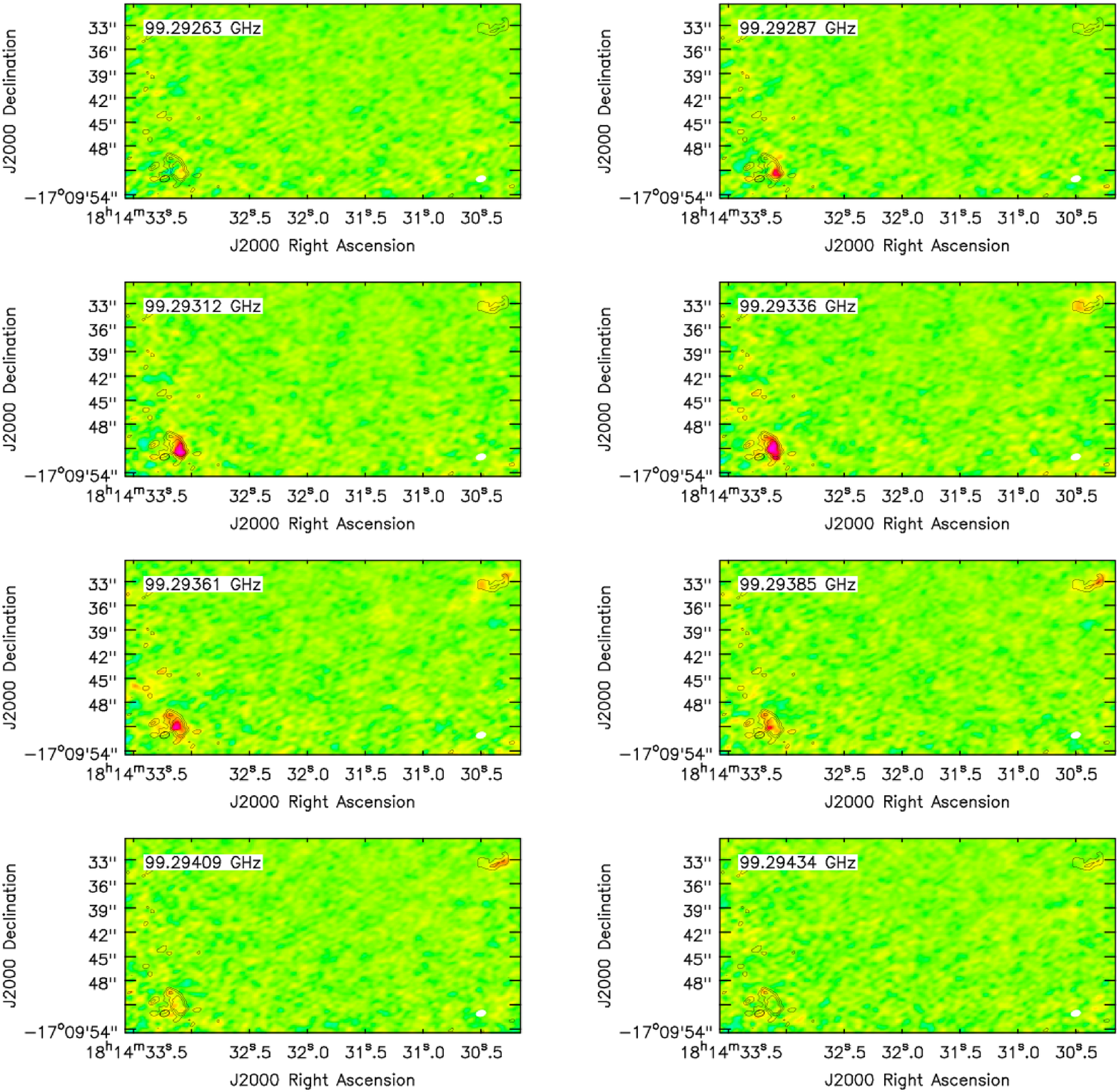}
\caption{Detail of regions showing line emission in the field of view of \gx: channel map at the SE and SW positions for the line centred at 99.2933 GHz. The contours show the integrated intensity after primary beam correction within the frequency range indicated in the first column of Table~\ref{tab:lineloggx}.}
\label{fig:line_maps2}
\end{figure*}

In contrast, we did not detect line emission at the position of any source down to 3\,$\sigma$ limits of 
0.65\,--3.7~K~km~s$^{-1}$ (see Table~\ref{tab:linelog}). However, we did detect line emission at three different positions within the field of view of \gx\ (see Table~\ref{tab:lineloggx} and Fig.~\ref{fig:gx13_composite}). This is the only source in the sample that shows line emission in its surroundings. Interestingly, we did not detect continuum emission down to noise levels of 30~$\mu$Jy/beam in any of the positions that show significant line emission. In addition, the lines have different spatial properties since one is spatially unresolved and the others are spatially resolved at scales of up to $\sim$6\arcsec\ (see Figs.~\ref{fig:line_images}--\ref{fig:line_maps2}).

\begin{table}
\begin{center}
\caption[]{Flux densities of the continuum emission at the positions of the observed XRBs and spectral index, $\alpha$, between 85.764 and 97.878~GHz, where $F_\nu \propto \nu^{\alpha}$.}
\begin{scriptsize}
\begin{tabular}{lllllll}
\hline \noalign {\smallskip}
Source &  Peak flux & Spectral index \\
& (mJy/beam) & \\
\hline \noalign {\smallskip}
\cirx\ & 
13.69\,$\pm$\,0.05 & -0.09 $\pm$0.01\\
\noalign {\smallskip}   
\scox\  & 
2.48\,$\pm$\,0.03 & 
   -0.007$\pm$0.008 \\
\noalign {\smallskip} 
\grs\ & 3.63\,$\pm$\,0.04 & 
0.015$\pm$0.008 \\
\noalign {\smallskip}
\gx\ & 
 0.75\,$\pm$\,0.02 
 & 0.005$\pm$0.005\\
\noalign {\smallskip}
\cygx\ & 5.69\,$\pm$\,0.05 & -0.02\,$\pm$\,0.01 \\
\noalign {\smallskip} 
\label{tab:fluxlog}
\end{tabular}
\end{scriptsize}
\end{center} 
\end{table}

\begin{table}
\begin{center}
\caption[]{3\,$\sigma$ upper limits to line intensities at the position of the XRBs in the data cubes for the lower and upper side bands, with central frequencies of 85.764 and 97.878~GHz, respectively.}
\begin{scriptsize}
\begin{tabular}{lll}
\hline \noalign {\smallskip}
Source & \multicolumn{2}{c}{Line intensity (K~km~s$^{-1}$)} \\
& Lower Side Band & Upper Side Band \\
\hline \noalign {\smallskip}
\cirx\ & $<$3.3 & $<$3.7  \\
\noalign {\smallskip}   
\scox\  & $<$2.1 & $<$2.4 \\
\noalign {\smallskip}   
\grs\ & $<$1.5 & $<$1.5 \\
\noalign {\smallskip}   
\gx\  & $<$0.65 & $<$0.65 \\
\noalign {\smallskip}   
\cygx\  & $<$1.0 & $<$1.0\\
\noalign {\smallskip} 
\label{tab:linelog}
\end{tabular}
\end{scriptsize}
\end{center} 
\end{table}

\begin{table*}
\begin{center}
\caption[]{Properties of the lines detected in the field of view of \gx. The positions and peak intensities have been derived from fitting the emission region with a Gaussian in the integrated intensity map after primary beam correction and within the frequency range indicated in the first column. 
The last  column indicates the peak frequency of the emission line fitted in the spatial region derived from the spatially integrated emission.
}
\begin{scriptsize}
\begin{tabular}{lllllllll}
\hline \noalign {\smallskip}
Frequency range & \multicolumn{2}{c}{Position} & Peak temperature & Peak frequency \\
(GHz) & RA (J2000) & Dec (J2000) & (K~km~s$^{-1}$) & (GHz)\\
\hline \noalign {\smallskip}
86.2344--86.2369 & 18 14 30.35 & -17 09 01.0 & 340.8\,$\pm$\,1.5 & 86.2362 \\ 
\noalign {\smallskip}
96.7326--96.7333 & 18 14 33.13 & -17 09 50.9 & 8.3\,$\pm$\,1.0 & 96.7329 \\
 \noalign {\smallskip}  
99.2928--99.2940 & 18 14 33.12 & -17 09 50.9 & 10.4\,$\pm$\,1.5 & 99.2933 \\
\noalign {\smallskip}
99.2928--99.2940 & 18 14 30.38 & -17 09 33.3 & 5.0\,$\pm$\,0.7 & 99.2938 \\
\noalign {\smallskip} 
\noalign {\smallskip} \hline 
\label{tab:lineloggx}
\end{tabular}
\end{scriptsize}
\end{center} 
\end{table*}


\section{Discussion}
\label{sec:discussion}

Radio continuum emission at centimetre wavelengths was previously detected for all the XRBs in our sample and was identified with a jet origin. Emission up to at least $\sim$40~GHz has also been observed for all the sources except \gx, for which reported detections span up to approximately 10~GHz. Given the previous measurements at lower frequencies and the spectral index of the emission at each accretion state of each source, the flux densities reported here are not unexpected. We give a detailed comparison for each source in Sect.~\ref{sec:cont}. 

We also searched for line emission down to a spectral resolution of $\approxlt$\,1~km s$^{-1}$ at the positions of the XRBs and their surroundings. Water maser emission has been observed in the accretion discs of supermassive BHs \citep[e.g.][]{humphreys13apj} and at the regions where their jets impact molecular clouds \citep[e.g.][]{peck03apj}. Given the similarities between accretion processes in XRBs and supermassive BHs, we would expect line emission to also be present in XRBs. While the upper limits to the presence of lines in our observations are quite stringent, all the observations showed continuum emission from the jet, which at the same time may hinder the detection of disc lines. Alternatively, strong line emission may only become significant over the continuum in the presence of flares. For example, \citet{1915:eikenberry98apjl} identified the NIR He$_{I}$ and Brackett $\gamma$ emission lines in \grs\ as arising in the accretion disc and found that their flux varied proportionally with the IR continuum flux, indicating that the lines were radiatively pumped by the flares. In this sense, our observations show relatively weak jet emission compared to the radio flares observed up to Jansky levels in these sources during other periods, such as changes of accretion state. For comparison, the Brackett $\gamma$ emission line reported by \citet{1915:eikenberry98apjl} is only $\sim$\,5~mJy above the continuum when the continuum is 40~mJy, but it brightens to $\sim$\,10 mJy above the continuum when the continuum increases up to 60~mJy (see their Fig.~1). 

As opposed to the positions of the XRBs, continuum emission is unlikely to be the reason for not detecting line emission near the XRBs and at the regions close to \cirx\ and \cygx,\ where evidence for jet-ISM interaction had previously been observed. A potential reason for the lack of lines is that the lines are weak or extend over large regions. For comparison, the molecular lines of SiO and CH$_3$OH reported by \citet{1915:tetarenko18mnras,tetarenko20mnras} and traced back to jet-ISM interactions have temperatures generally below 0.6~K~km~s$^{-1}$ and are extended over relatively large regions, at near-arcminute scales.\  Therefore, they would not be detected in our observations, which reached sensitivities of 0.6--4~K~km~s$^{-1}$ (see Table~\ref{tab:linelog}) and could only recover angular scales up to $\sim$15\arcsec. 
Another potential reason could be that 
the medium that the jet or the wind is propagating through has a low density compared to the average ISM density, $\sim$1~cm$^{-3}$. Based on the observation of \grs\ jet ejections travelling at a constant speed to distances of at least $\sim$0.04~pc \citep[][]{1915:mirabel94nature,fender99mnras,1915:miller-jones07mnras} and on jet energetic arguments, \citet{heinz02aa} estimated that the density around this XRB was $\sim$10$^{-3}$~cm$^{-3}$ (unless the jets were narrower than 0.15\deg\ or extremely energetic,  10$^{47}$~ergs). They suggested that the region around \grs\ was occupied by the hot ISM phase or that previous activity of the jets (or the supernova at which the progenitor formed) could have created an evacuated bubble around it. Interestingly, \citet{haid18mnras} found that, in the warm ionised ISM, stellar winds dominate the momentum injection when compared to ionising radiation for massive stars. Therefore, in LMXBs, accretion disc winds could provide an alternative to stellar winds for creating a cavity around the XRB or as a source of momentum injection in a previously evacuated bubble. A dedicated study considering the ratio of wind to XRB ionising radiation is needed to advance further on this topic. 

Finally, while the shocked region around \cirx, between 20 and 50\arcsec\ from the source (corresponding to $\sim$0.9--2.3~pc at the distance of the source), was found to be bright in diffuse X-rays and radio emission \citep[see their Fig.~2]{cirx1:sell10apj}, we did not detect any continuum emission in the same areas down to noise levels of 45~$\mu$Jy/beam (but we note that we cannot exclude emission at the low level of $\sim$24~$\mu$Jy/beam that was detected in the previous radio observations at 5.5~GHz). 
In contrast to \cirx, the shocked emission region around \cygx\ is bright in H$\alpha$ and O$_{III}$ emission and in 1.4~GHz radio continuum emission and is located at the rim of a jet-inflated nebula of $\sim$5~pc in diameter \citep{cygx1:gallo05nature,cygx1:russell07mnras}. However, our continuum maps again show no significant emission (noise level of  40~$\mu$Jy/beam) in the region where the ring in radio emission at 1.4~GHz was previously observed at a flux level of 0.2~mJy/beam. A potential reason for the non-detection of the continuum in this case may be a spatial extent larger than that recoverable with our observations, $\sim$15\arcsec\ (0.16~pc at the distance of \cygx). 

In summary, we found neither continuum nor line emission that could signal the presence of shocks at regions near the probed XRBs, including regions around \cirx\ and \cygx\ where evidence for jet-ISM interaction had been previously observed.\ However, we also note that many of the molecular tracers of jet-ISM interactions, such as the shock tracer CS or the high-density tracers HNCO or HCO+, remain unexplored for the sources of this study. Nevertheless, we discovered three different line emission features in the field of view of \gx\ that could signal the presence of a yet unknown star-forming region, albeit one probably unrelated to \gx. We discuss these features in detail in Sect.~\ref{sec:linegx}.

\subsection{Continuum emission from a jet}
\label{sec:cont}
\subsubsection{\cirx}
Using the ephemeris in \citet{cirx1:nicolson07atel} derived from the onset times of 21 well-observed radio flares spanning the past 28 years, the ALMA observation occurred at phase $\sim$0.08. This is $\sim$6 days later than the X-ray peak detected in the Monitor of All-Sky X-ray Image (MAXI) light curves between Modified Julian Date (MJD) 57675.5 and 57677.5,  and it is consistent with previous observations that indicated that the radio flares occur later in the orbit compared to the X-ray flares.

We detected significant continuum emission with a peak flux of 13.69\,$\pm$\,0.05 mJy/beam and a spectral index of  --\,0.09\,$\pm$\,0.01 between 84 and 99 GHz. The emission is unresolved, implying a region of emission smaller than $\sim$10$^{17}$\,cm for the reported beam of 0.67\arcsec\,$\times$\,0.63\arcsec\ and the distance of the source. In observations performed with the Australian Telescope Compact Array (ATCA) at a resolution of 0.44\arcsec\,$\times$\,0.23\arcsec\ and frequencies of 33 and 35~GHz, extended emission was reported on two different occasions \citep{cirx1:calvelo12mnras,cirx1:coriat19mnras}.  
The flux in our observations is stronger than that previously reported at 33--35~GHz at the core ($\sim$7 and 0.7 mJy/beam). If the extension of the emission had been similar to that reported at 33-35~GHz, we would have detected it despite our slightly poorer resolution. Thus, we conclude that our non-detection of extended emission is more likely related to the higher frequency of our observations since the extension of the emission significantly decreases as the frequency increases \citep{cirx1:coriat19mnras}. 

\begin{figure}[ht]
\includegraphics[angle=0.0,width=0.45\textwidth]{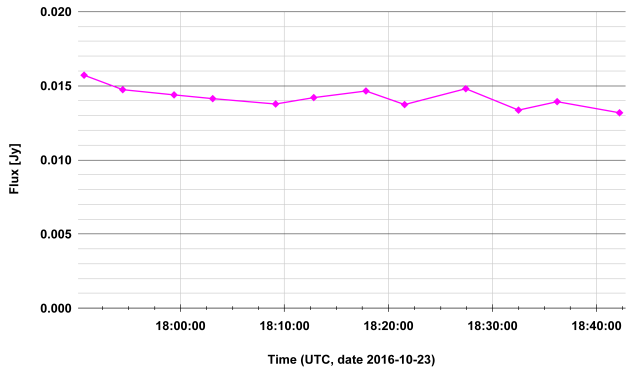}
\caption{Light curve of \cirx\ during the ALMA observation. The errors are $\sim$ 2\% for each point (the size of the symbols).}
\label{fig:cirx1lc}
\end{figure}

This source is the only one in our sample that shows a negative spectral index.\ This indicates optically thin emission, and it is consistent both with previous radio measurements at similar orbital phases that were interpreted as the decline of a periastron passage flare \citep{cirx1:calvelo12mnras} and with the flux decay observed in the light curve of the ALMA observation (see Fig.~\ref{fig:cirx1lc}).

\subsubsection{\scox}

We detected unresolved emission with a flux density of 2.48~mJy/beam and a flat spectral index between 84 and 99 GHz at the position of \scox. Previous emission from lobes moving away from the core, such as those detected by \citet{scox1:fomalont01apj}, occurred at milliarcsecond scales (1 mas corresponding to 2.8 AU at the distance of \scox) and would not have been resolved at the resolution of our images (see Table~\ref{tab:obslog2}). However, the flat spectral index indicates that we are most likely detecting core-only (or core-dominated) emission since the lobe emission always had negative spectral indices,
as expected from optically thin emission \citep{scox1:fomalont01apj}. Based on a timing analysis of contemporaneous X-ray observations, \citet{scox1:motta19mnras} associated the launch of the ultra-relativistic outflows causing the lobe emission with particular accretion states (between the so-called `normal' and `horizontal' branches). We produced X-ray colour-colour diagrams with data from MAXI (see Appendix \ref{sec:app1}) and found that, at the time of the ALMA observations, the source was instead at the so-called `flaring branch', a state associated with the weakest and least variable radio emission \citep{scox1:hjellming90apj}. The flux level detected by ALMA is of the order of the fluxes detected at such a state by \citet{scox1:hjellming90apj} between 1 and 8~GHz and with flat spectral index, although they also detected additional components during the same state with a negative spectral index. In conclusion, our detection is consistent with the jet emission extending from centimetre to millimetre wavelengths with a flat spectral index during the flaring branch, although simultaneous measurements at both wavelengths are needed to confirm this.

\subsubsection{\grs}

We detected spatially unresolved emission with a flux density of 3.63~mJy/beam and a slightly inverted spectral index between 84 and 99 GHz at the position of \grs. Based on MAXI colour-colour diagrams, we infer that, at the time of the ALMA observations (see Appendix \ref{sec:app1}), the source was most likely in the `A' or `B' state, according to the classification of \citet{1915:belloni00aa}, characterised by a substantial contribution of a disc component to the X-ray spectrum and low-level radio emission. \citet{1915:klein-wolt02mnras} associated the radio emission during those states with either a hard component or a relic from a previous large radio event. The low flux density of the ALMA observation within this accretion state is consistent with a relatively radio-quiet period during that state.

\subsubsection{\gx}

We detected, for the first time, compact continuum emission at a frequency of 92~GHz from \gx. The flux density at centimetre wavelengths ranges from $\sim$0.25~mJy/beam to $\sim$1.3--7.2~mJy/beam and is correlated with the accretion state of the source as determined from X-ray observations \citep{gx13:garcia88apj,gx13:homan04aa}. The ALMA observations indicate a flux of 0.75\,$\pm$\,0.02 mJy/beam, in between the two ranges. The spectral index is flat, similar to that observed at radio frequencies during states with weak radio emission. 

\subsubsection{\cygx}

The second observation of \cygx, based on hardness-intensity diagrams extracted from MAXI data (see Appendix \ref{sec:app1}), was performed during a spectrally hard state. While the MAXI light curves do not cover the date of the first observation, we conclude, based on the similarity of the hardness and intensity before and after that observation, that the accretion state most likely did not change between the two observations. 
Our detection of continuum emission with a flat spectral index is consistent with previous findings of a flat spectra from radio to millimetre wavelengths and attributed to a compact jet \citep{cygx1:fender00mnras}. 

\subsection{Line emission in the field of view of \gx}
\label{sec:linegx}
Similarly to all the other observed sources, no spectral lines were observed at the position of \gx. In contrast, we detected three different line emission features in its field of view. 

The strongest feature was observed 27\arcsec\  north-west (NW) of \gx\ at a frequency of 86.2361~GHz  (see Table~\ref{tab:lineloggx} and Fig.~\ref{fig:line_images}). We tentatively identified the line with emission from vibrationally excited SiO (v=1, J=2-1) with a redshift of 25 km s$^{-1}$. The emission is spatially unresolved, implying an emission size of less than 8\,$\times$\,10$^{16}$~cm assuming the source is at a similar distance as \gx\ (i.e. 7~kpc). Assuming the line identification is correct and that its shift is not intrinsic to the source but is instead due to its position in the galaxy, we obtain 2.5 and 13.3~kpc as near and far kinematic distances, implying emission sizes of less than $\sim$5\,$\times$\,10$^{16}$~cm and $\sim$3\,$\times$\,10$^{17}$~cm, respectively.

We found a source coincident with the position of the line in the galactic young stellar object (YSO) catalogue of \citet{marton19cat}, which was produced by running a supervised machine learning algorithm on the combined  data from the Gaia DR2 database, the Wide-field Infrared Survey Explorer (WISE), and Planck. 
The source coincident with the SiO emission, J181430.35-170901.0, has a 57\% and 42\% probability of being a YSO or an evolved star, respectively, when all the WISE bands are used, and a 43\%\ and 55\% probability if WISE bands W3 and W4 are neglected. SiO masers are uncommon in star-forming regions \citep{zapata09apj}, with only six reported detections in YSOs so far \citep{cho16apj}. Additionally, if the SiO maser is associated with a star-forming region, we would expect to detect continuum emission in the ALMA maps unless the region is quite distant. For reference, a one-solar-mass core would have a flux of 0.3~mJy if placed at the near kinematic distance of 2.5~kpc, thus easily detectable in our maps (for this estimate we considered a $\sim$20~K dust temperature, the typically adopted dust opacity, and a gas-to-dust mass ratio of 100, and we assumed that 
the emission at 90~GHz was optically thin). Conversely, assuming the source of the SiO to be an evolved star at a distance of 2.5~kpc, we estimate, following calculations by \citet{muller07apj,muller08apj}, a stellar flux of $\sim$25\,$\mu$Jy/beam and a dust envelope flux of $\sim$65\,$\mu$Jy/beam for a mass loss rate of $\approxgt$6\,$\times $10$^{-6}~\Msun$ at 90~GHz. Therefore, our 3$\sigma$~upper limit of 90~$\mu$Jy/beam indicates that our maps are not sensitive enough to detect the combined flux of the star and the dust envelope of a typical evolved star such as W Hya, for which a mass loss rate of 4.9\,$\times$\,10$^{-7}~\Msun$ within a radius of about 70 AU has been derived \citep{vlemmings11apj}. Consequently, an evolved star origin for the SiO line remains a viable solution.
Alternatively, if the source is relatively far ($\approxgt$4~kpc), both a YSO and an evolved star remain possibilities. Further observations of the detected line and of other SiO transitions at a higher angular resolution are necessary to confirm the maser nature of the emission and the line identification.

In addition to the potential SiO maser NW of \gx, we also detected spatially extended line emission with a frequency of $\sim$99.2935~GHz at 12\arcsec\ south-west (SW) and 38\arcsec\ south-east (SE) of \gx. If associated with H(40) $\alpha$, the line has a blueshift of 820~km~s$^{-1}$ or, if associated with SO, a redshift of 18~km~s$^{-1}$. At the SE position, we detected a second line with a frequency of 96.7329~GHz. If associated with H$_2$O, the line has a blueshift of 725\,km~s$^{-1}$ or, if associated, with CH$_3$OH, a redshift of 20~km~s$^{-1}$. The spatial coincidence of two lines at the SE position indicates that a similar redshift of 18--20~km~s$^{-1}$ for the lines, and thus an identification with SO and CH$_3$OH, is more plausible. The detection of the same line in the SE and SW positions may indicate that both regions are also related. 

\begin{figure*}[ht]
\includegraphics[angle=0.0,width=1.0\textwidth]{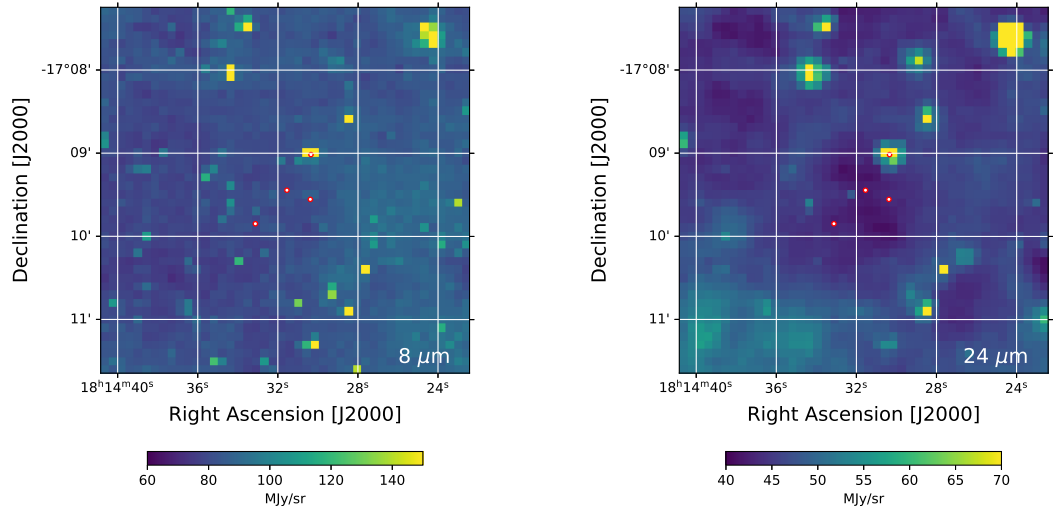}
\vspace{-0.5cm}
\caption{$Spitzer$ GLIMPSE (8~$\mu$m) and MIPSGAL (24~$\mu$m) maps centred at the position of \gx. The red dots mark the positions of \gx\ and the positions of the lines detected in the ALMA maps.
}
\label{fig:mipsgal}
\end{figure*}

\begin{figure*}[ht]
\hspace{0.5cm}
\includegraphics[angle=0.0,width=1.02\textwidth]{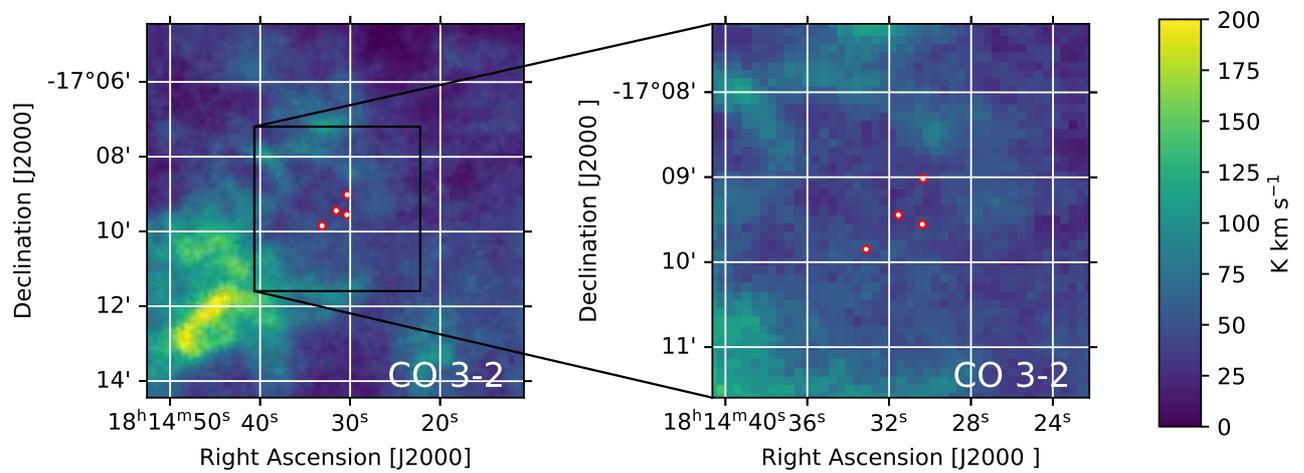}
\vspace{-1cm}
\caption{CO (3-2) maps taken with the JCMT and centred at the position of \gx. The red dots mark the positions of \gx\ and the positions of the lines detected in the ALMA maps.}
\label{fig:jcmt}
\end{figure*}

Given the uncertain origin of the SiO emission (YSO or evolved star) and the occurrence of CH$_3$OH and SO molecules in regions of star formation \citep[e.g.][]{oberg14fa,li15apjl,aikawa20apj}, we searched for the signature of a parent molecular cloud in archival millimetre single-dish images that would indicate the existence of such a star-forming region, thus linking the three line emission regions. None of the positions were detected in the ATLASGAL survey at 850~$\mu$m with a noise level of 50 mJy/beam (but we note the coarse resolution of this survey, 20\arcsec) nor in the James Clerk Maxwell Telescope (JCMT) continuum maps at the same wavelength. This excludes the existence of at least an OB star cluster in this region and is consistent with the absence of any continuum detection in our images down to an rms of 30~$\mu$Jy/beam. We also inspected the images of the {\it Spitzer} MIPSGAL 24~$\mu$m \citep{carey09pasp} and GLIMPSE 8~$\mu$m \citep{churchwell09pasp} surveys (see Fig.~\ref{fig:mipsgal}). The source coincident with the SiO emission, J181430.35-170901.0, is clearly visible at micrometre wavelengths in both images. It is also apparent that the positions of \gx\ and the line detections SE and SW of \gx\ are instead located in a relatively dark area in the MIPSGAL 24~$\mu$m image, suggesting that no dust is present in that region. This is consistent with the absence of significant CO (3-2) emission in the JCMT/HARPS 345~GHz map (project code:  M10AU20) of the same region (see Fig.~\ref{fig:jcmt}), although we note that the cavity, or `dark' region, surrounding \gx\ in the MIPSGAL image is significantly smaller than that of the CO (3-2) map. The existence of such a dark region around \gx\ could suggest a link between the XRB and the line emission regions, but this possibility remains speculative at this stage. Finally, we note that an old 20~cm image of the region \citep[][see their Fig.~3]{gx13:garcia88apj} seems to show emission at the position of \gx\ and the SE line emission region at a level of $\sim$1~mJy but no emission at the NW or SW positions.\ Additionally, a 6~cm Very Large Array (VLA) image in August 1999 covering the region of the ALMA observation down to an rms of 50~$\mu$Jy/beam only shows significant radio emission at the position of \gx. 

In summary, the field of view of \gx\ shows a set of features of which the strongest (SiO) is best interpreted in the context of an evolved star or a star-forming region that is currently unknown or has not been detected in continuum emission to date. The origin of the other features remains unexplained. 

\section{Summary and conclusions}

We observed five XRBs at millimetre wavelengths with ALMA with the goal of exploring the existence of atomic and molecular line emission from their accretion discs, both close to the XRBs, where powerful X-ray and optical winds are expelled, and farther away, where sites of jet-ISM interaction are known and shocks are expected. All the sources show continuum emission consistent with a jet origin and previous observations below 40~GHz. In contrast, our exploratory search for line emission only resulted in upper limits at the position of all sources and at the sites near \cirx\ and \cygx,\ where previous signatures of jet-ISM interaction had been detected. Further searches for disc lines could specifically target either accretion states during which no, or only weak, jet emission is detected or, alternatively, states during which bright flares could pump line emission. Instead, searches for lines in the XRB environments should be complemented with searches of line emission at arcminute angular scales and target additional lines.

Interestingly, the field of view of \gx\ shows a set of features whose origin remains enigmatic. While the strongest feature (SiO) can be interpreted as originating in an evolved star, it is better interpreted (if it is related to the other two line emission regions)  in the context of a star-forming region that is currently unknown or has not yet been detected in continuum emission. An association of these features with \gx\ remains speculative at this stage. Follow-up observations of CO at higher sensitivities could help us understand whether the sites of emission are related and shed light on the nature of this region. This should be complemented with observations of other transitions of the candidate molecules to confirm line identification.

\begin{acknowledgements}  
We thank the anonymous referee for a very useful report that helped to improve 
the quality of this paper. We also thank Mickael Coriat for providing the radio maps for
\cirx\ and Elena Gallo and Dave Russell for those of \cygx. MDT thanks Yuxin Lin for 
help in producing figure 9.
This paper makes use of the following ALMA data:
   ADS/JAO.ALMA\#2016.1.00496.S. ALMA is a partnership of ESO (representing
   its member states), NSF (USA) and NINS (Japan), together with NRC
   (Canada) and NSC and ASIAA (Taiwan), in cooperation with the Republic of
   Chile. The Joint ALMA Observatory is operated by ESO, AUI/NRAO and NAOJ.
\end{acknowledgements}


\bibliographystyle{aa}
\bibliography{masers}

\appendix

\section{Hardness-intensity diagrams}
\label{sec:app1}

\begin{figure*}
\includegraphics[angle=0.0,width=0.45\textwidth]{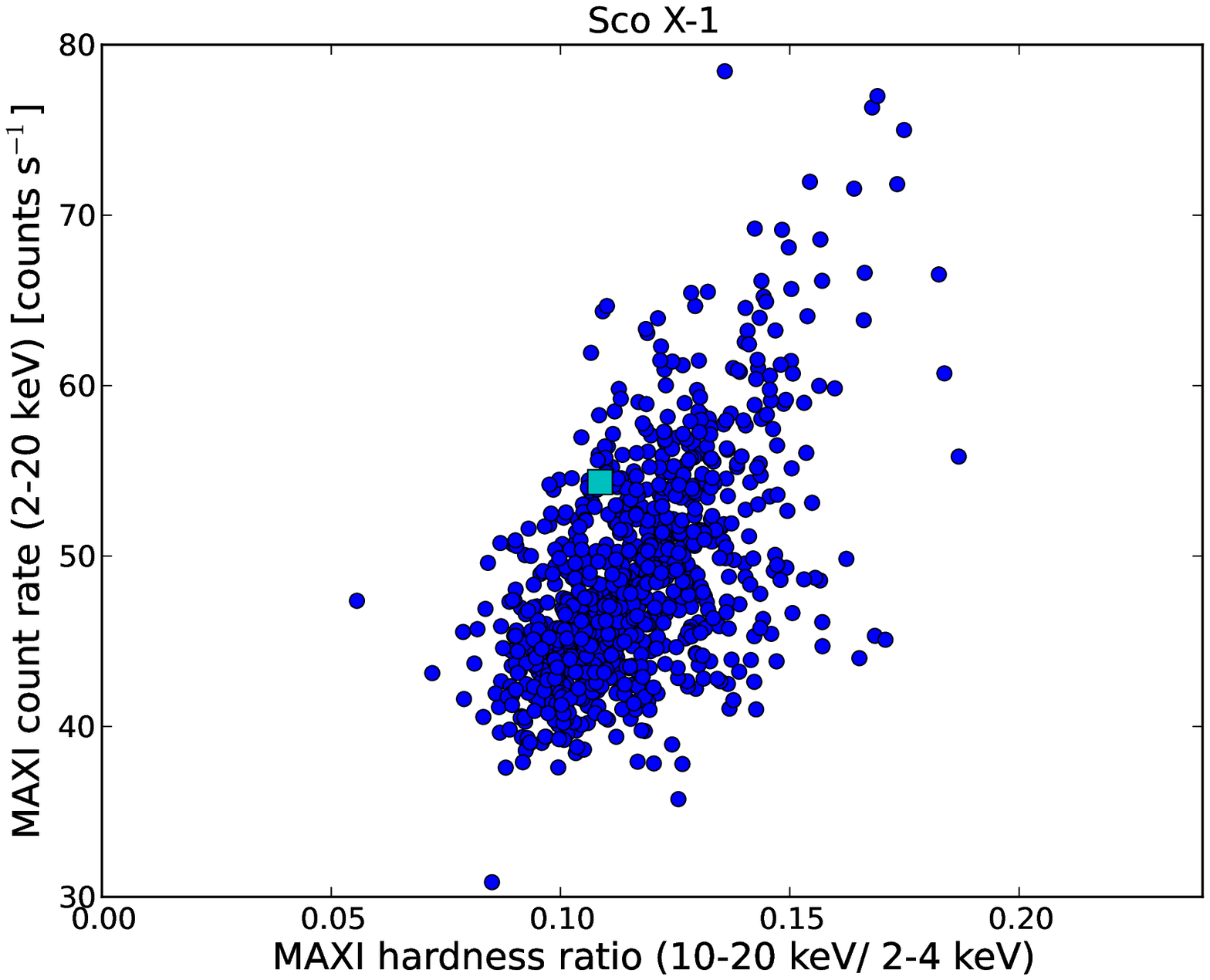}
\includegraphics[angle=0.0,width=0.45\textwidth]{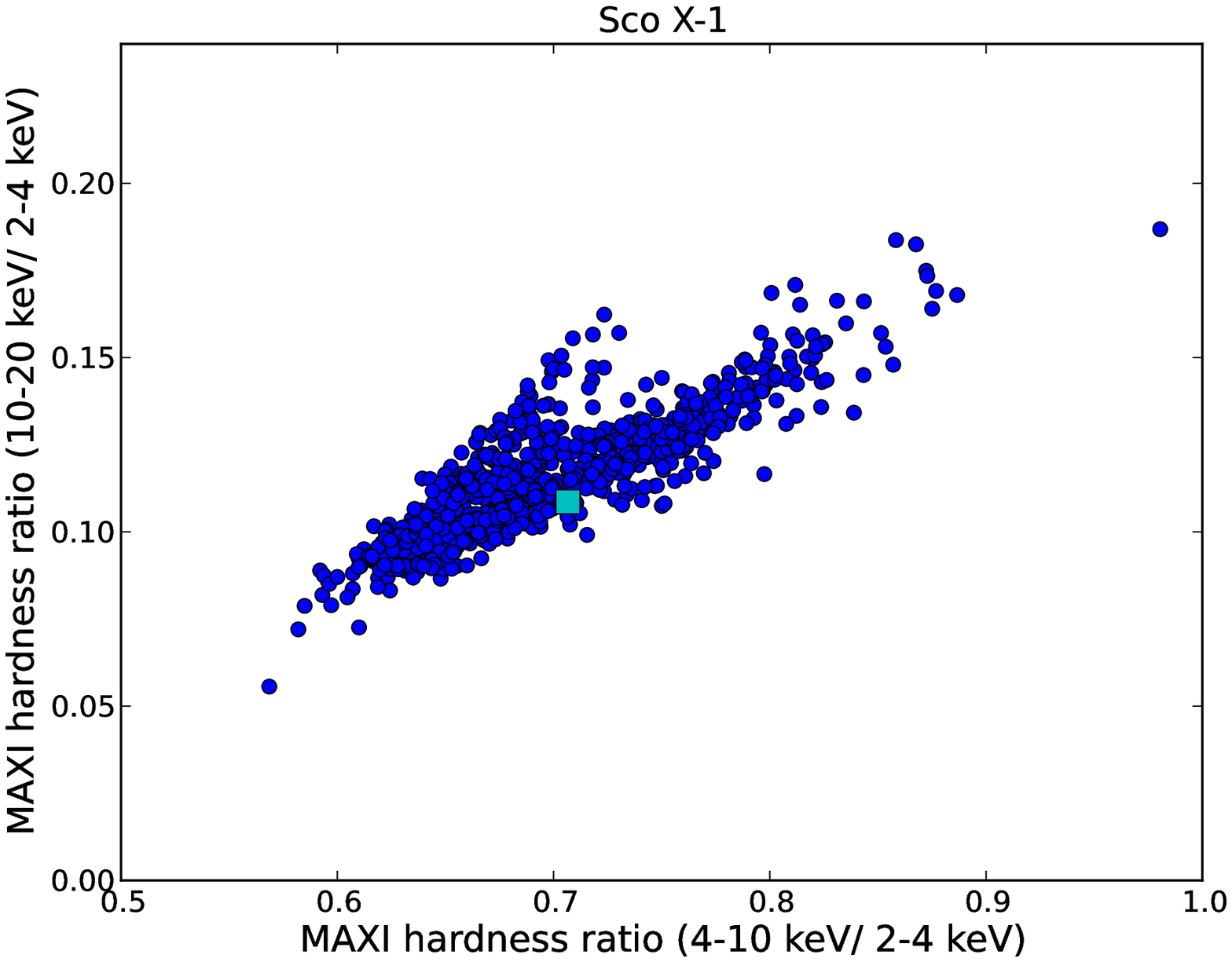}
\includegraphics[angle=0.0,width=0.45\textwidth]{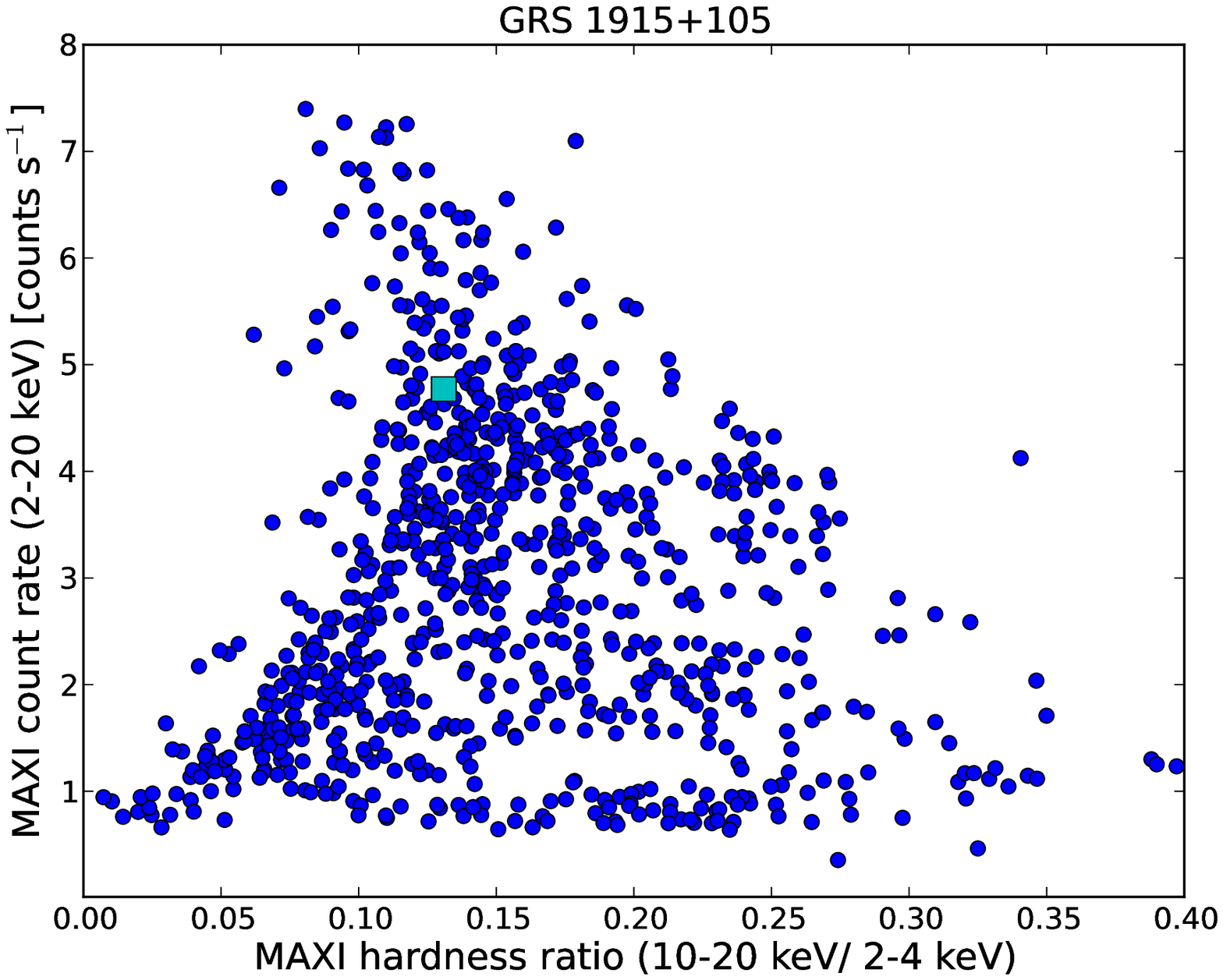}
\includegraphics[angle=0.0,width=0.45\textwidth]{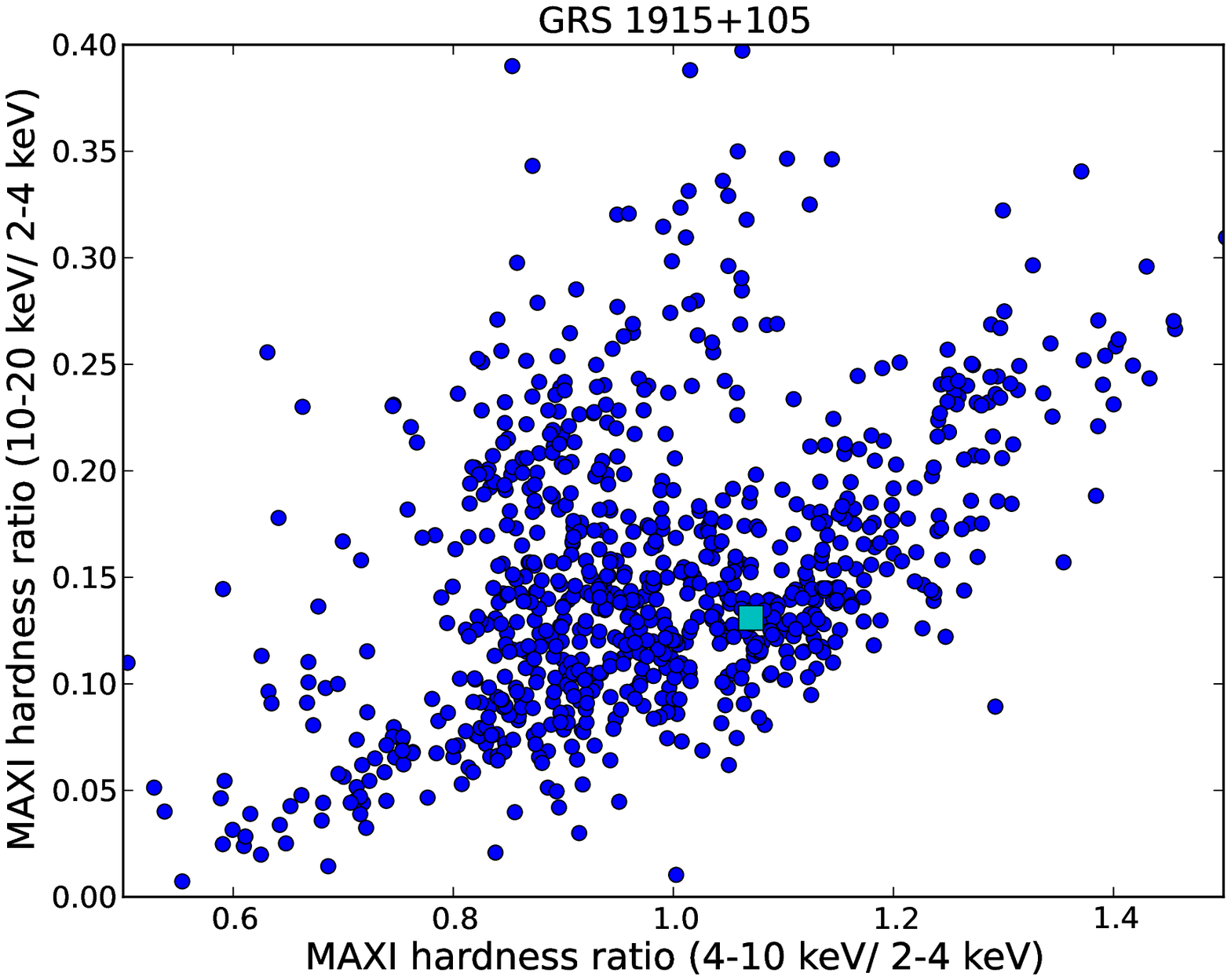}
\includegraphics[angle=0.0,width=0.45\textwidth]{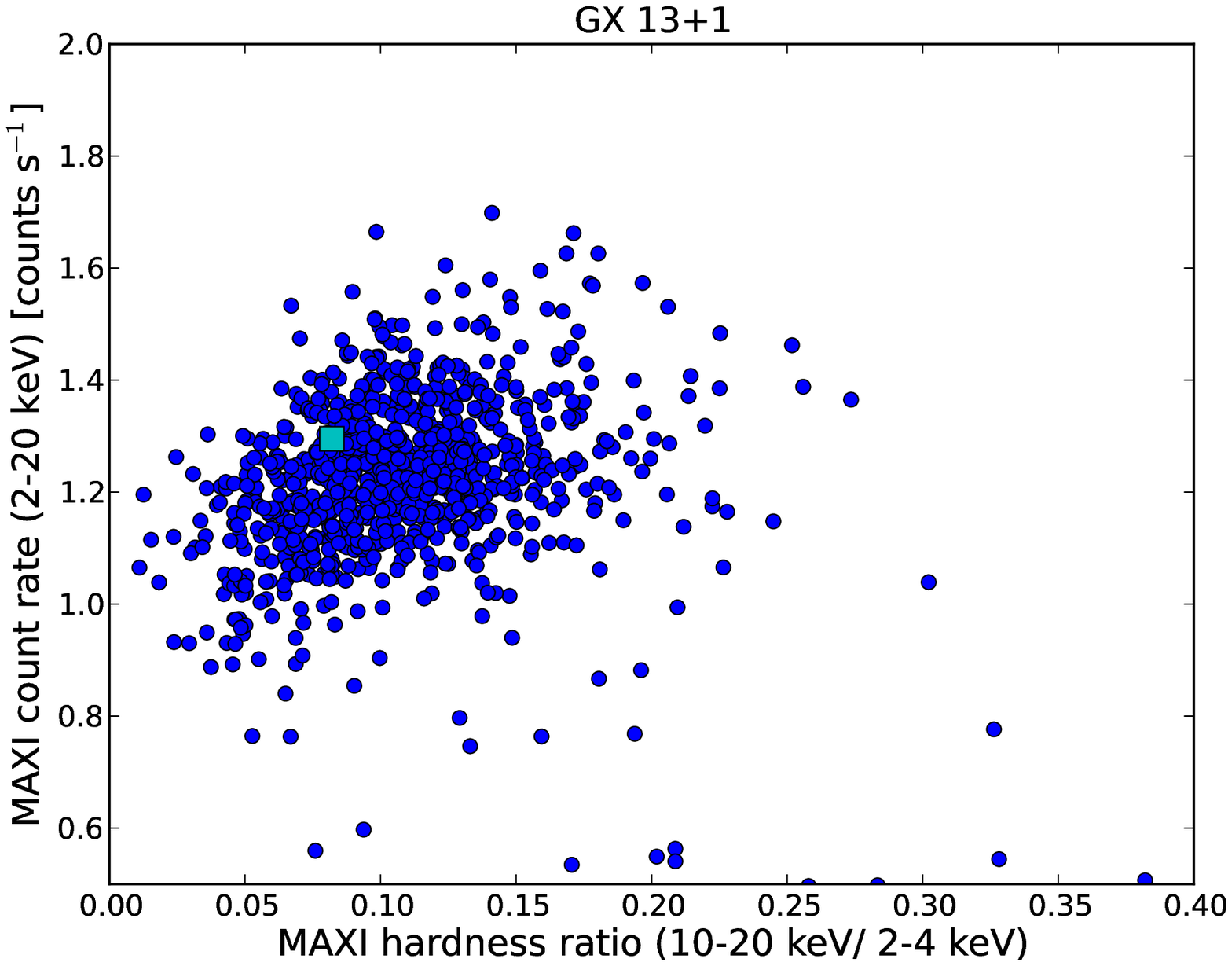}
\includegraphics[angle=0.0,width=0.45\textwidth]{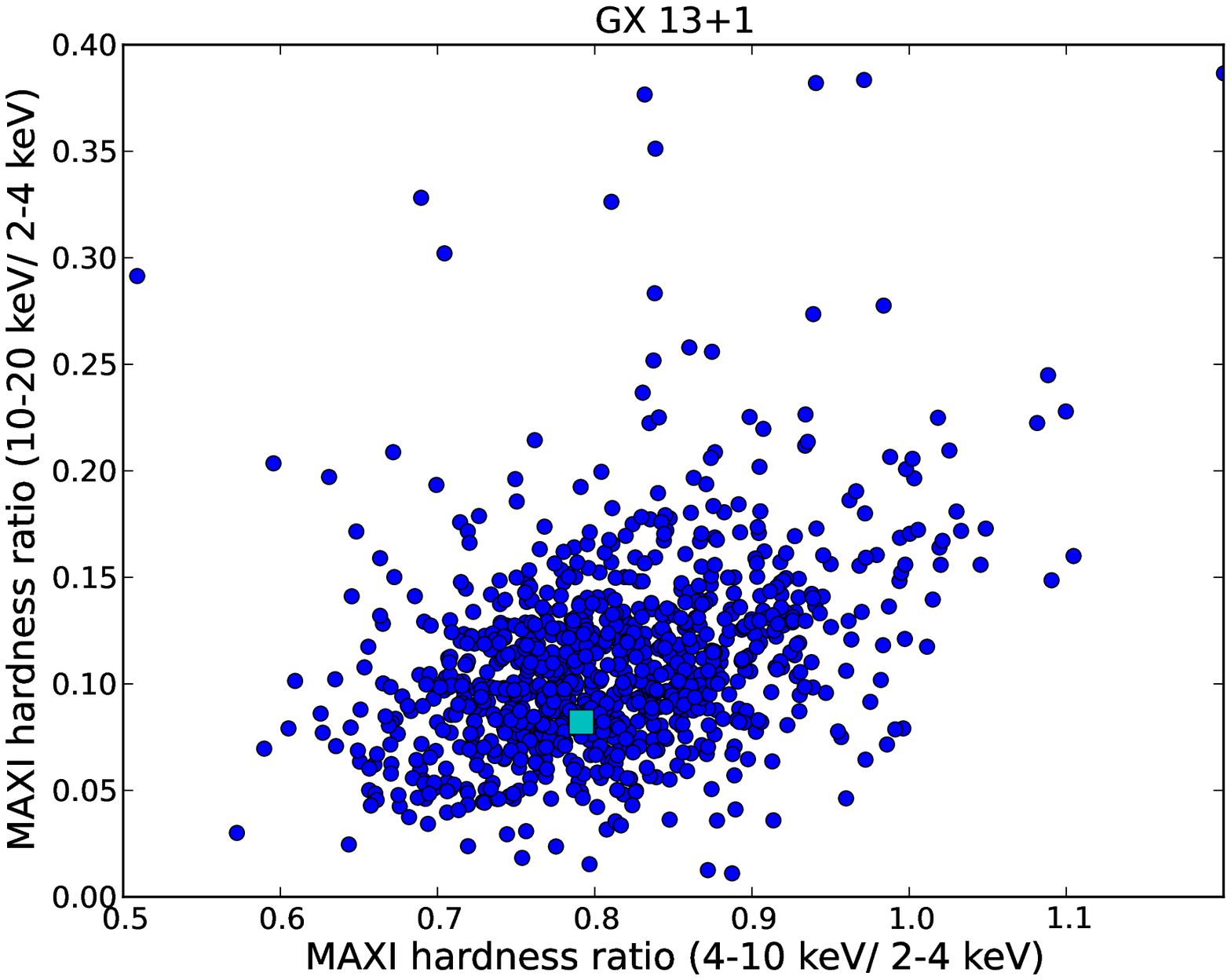}
\includegraphics[angle=0.0,width=0.45\textwidth]{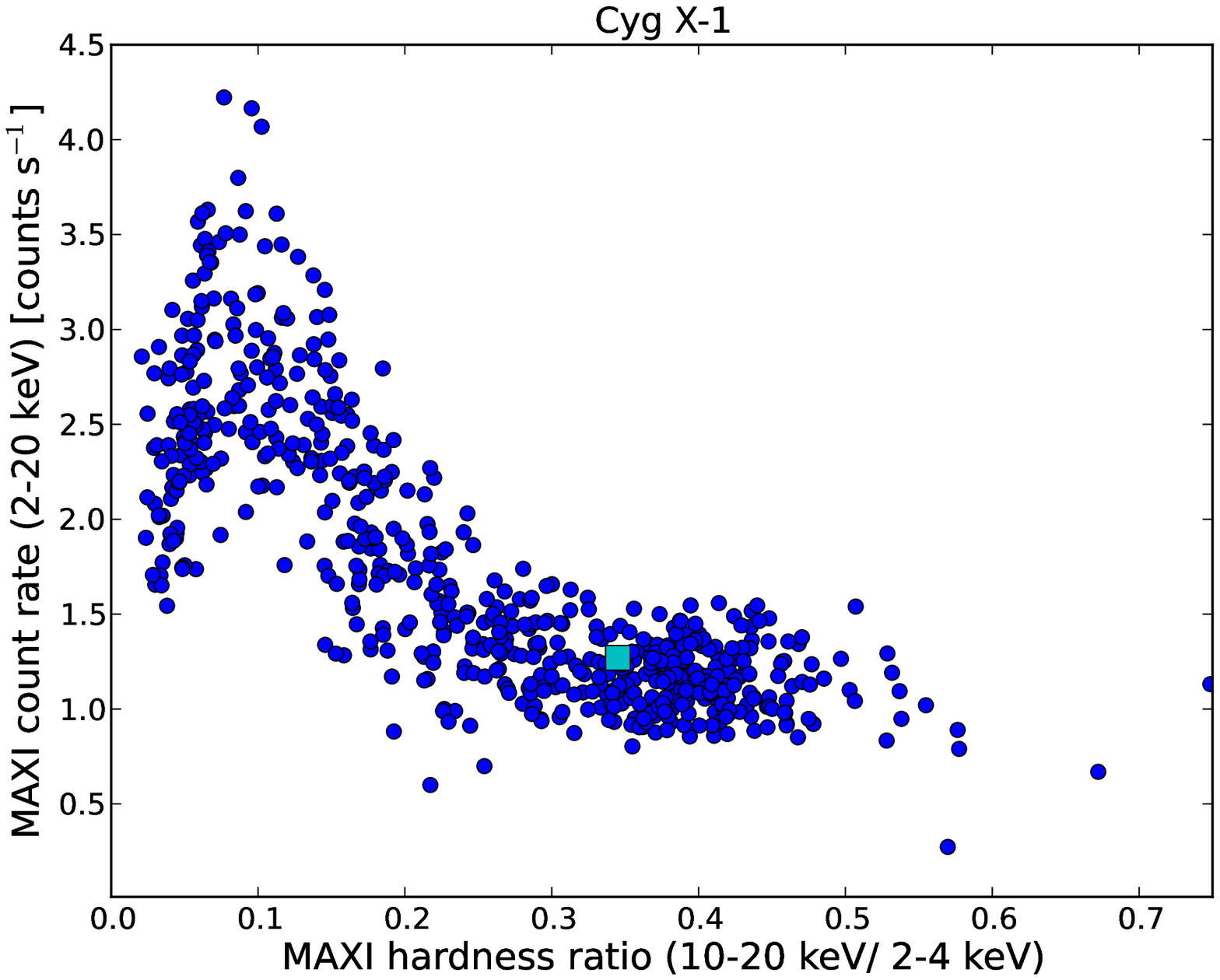}
\hspace{1.5cm}
\includegraphics[angle=0.0,width=0.45\textwidth]{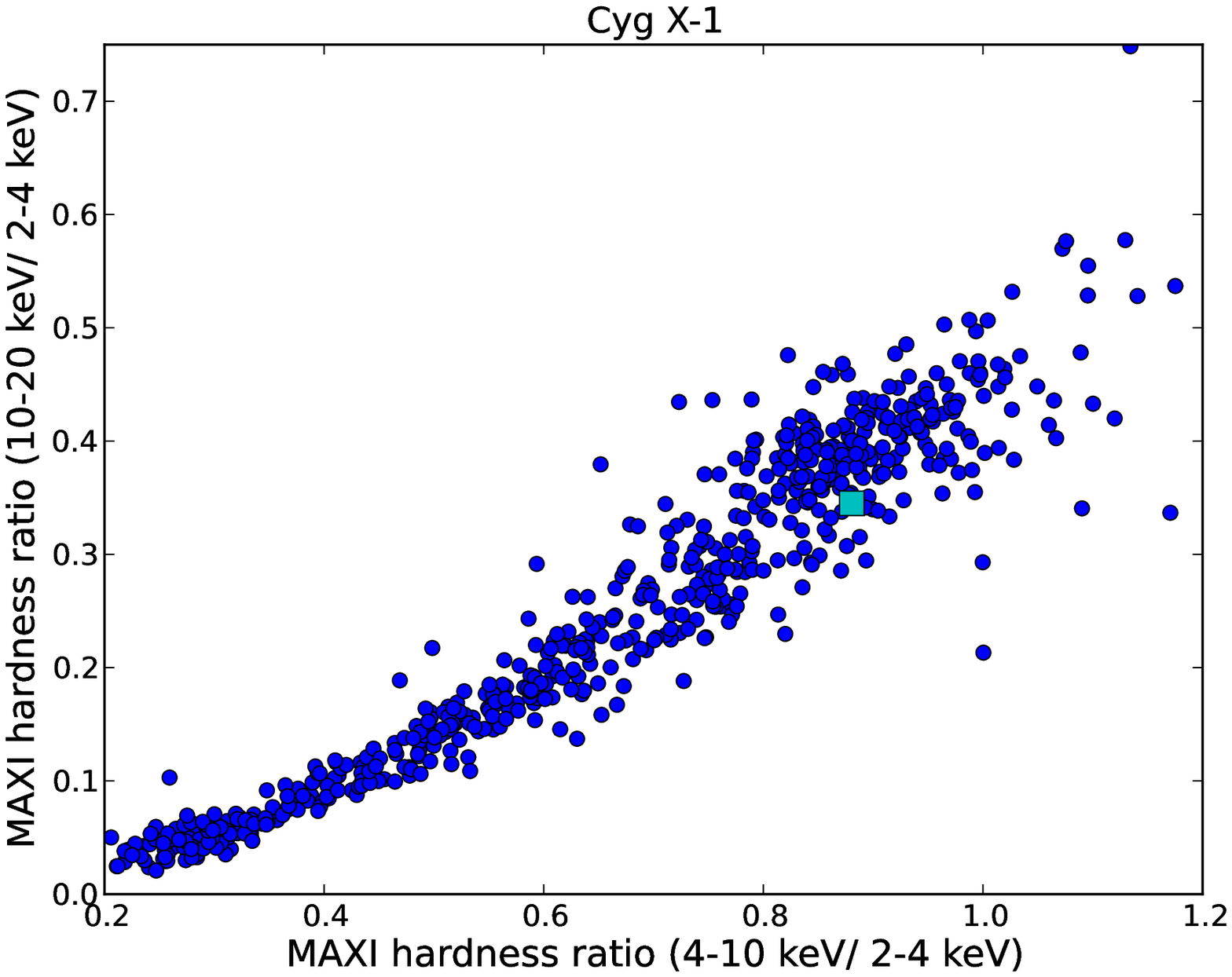}
\caption{Hardness-intensity (left) and colour-colour (right) diagrams based on MAXI data. Blue circles represent all MAXI daily measurements for each source in the shown energy range and between MJD 57000 and 58000 (as representative from the  MAXI lifetime). The cyan square point is the MAXI daily measurement at the time when the ALMA observation for each source was performed (see Table~\ref{tab:obslog}). For \cygx\ there were no MAXI data available on the day of the first observation, and consequently only the second observation is shown.}
\label{fig:maxihid}
\end{figure*}

Figure~\ref{fig:maxihid} shows the hardness-intensity and colour-colour diagrams from MAXI data used to determine the accretion states of \scox, \grs,\ and \cygx\ at the time of the ALMA observations. Based on the colour-colour diagrams \citep[and previous state classifications of Z sources based on these diagrams; e.g.][]{hasinger89aa}, we infer that the \scox\ observation was performed in the flaring branch. With only the MAXI data we cannot infer a state classification for \gx\ since no branches are distinguishable in the respective diagrams, as in, for example, \citet{homan16apj}. For \cygx\ we instead used the hardness-intensity diagram and followed the classification from \citet[][see their Fig.~7]{grinberg13aa} to infer that it was in a spectrally hard state. Finally, for \grs\ we used the colour-colour diagram and the classification from \citet[][see their Fig.~8]{1915:belloni00aa} and classify the ALMA observation as occurring during the A or B states.

\end{document}